\documentclass[10pt]{iopart}
\makeatletter
\renewenvironment{abstract}
  {\section*{\abstractname}}
  {}
\makeatother

\usepackage[backend=biber,language=auto,autolang=other, style=phys]{biblatex}  
\AtEveryBibitem{%
  \clearfield{language}%
  \clearlist{language}%
}
\usepackage{amssymb}
\usepackage{rotating}
\expandafter\let\csname equation*\endcsname\relax
\expandafter\let\csname endequation*\endcsname\relax
\usepackage{amsmath, amsbsy}
\usepackage{graphicx}
\usepackage[unicode=true, bookmarks=true, bookmarksnumbered=true, bookmarksopen=true, bookmarksopenlevel=1, breaklinks=false,pdfborder={0 0 0}, backref=false, colorlinks=true]{hyperref}
\hypersetup{citecolor=blue,linkcolor=blue}
\usepackage[dvipsnames]{xcolor} 
\usepackage{lipsum,multicol} 
\usepackage [english]{babel}
\usepackage [autostyle, english = american]{csquotes}
\MakeOuterQuote{"}
\usepackage{subcaption}

\usepackage{comment}
\usepackage{indentfirst}
\usepackage{adjustbox}
\usepackage{esint}

\begin{document}

\title[Improved n=1 Empirical Error Field Penetration Threshold Scaling]{Improved n=1 Empirical Error Field Penetration Threshold Scaling with Ohmic and L-Mode Conventional Tokamak Plasma Discharges}

\author{E.M. Bursch$^{1,2}$,
J.K. Park$^{3}$,
N.C. Logan$^{1,2}$,
F. Mao$^4$,
N. Wang$^5$,
C.F.B. Zimmermann$^{1,2}$,
R.J. Buttery$^6$,
C. Paz-Soldan$^{1,2}$,
M. Pharr$^{1,2}$,
L. Piron$^{7,8}$,
G. Szepesi$^9$,
H. Wang$^4$,
S.M. Yang$^{10}$,
JET Contributors$^{11}$,
EUROfusion Tokamak Exploitation Team$^{12}$
}
\address{
$^1$ Columbia University, New York, NY, United States of America \\
$^2$ Columbia Fusion Research Center, Columbia University, New York, NY, United States of America \\
$^3$ Seoul National University, Seoul, Republic Of Korea\\
$^4$ Institute of Plasma Physics Chinese Academy of Sciences, China\\
$^5$ Huazhong University of Science and Technology, Wuhan, China \\
$^6$ General Atomics, San Diego, CA, United States of America \\
$^7$ Consorzio RFX (CNR, ENEA, INFN, Universitá di Padova, Acciaierie Venete SpA) Padova, Italy \\
$^8$  University of Padova, Padova, Italy\\
$^9$ UKAEA, Culham Campus, Abingdon, United Kingdom\\
$^{10}$ Princeton Plasma Physics Laboratory, Princeton, NJ, United States of America\\
$^{11}$ See the author list in C.F. Maggi et al 2024 Nucl. Fusion 64 112012  \\ 
$^{12}$ See the author list of N. Vianello et al "Results from the last DD and DT JET campaigns in the framework of the EUROfusion Tokamak Exploitation Work Package Activity" 2026 submitted to Nuclear Fusion \\
 }

\ead{jkpark@snu.ac.kr}

\vspace{10pt}

\begin{abstract}
\noindent This paper presents an updated n=1 error field penetration threshold scaling, which increases fit quality compared to previous error field scaling laws, is produced from an expanded database, and exhibits reduced uncertainty in projections to future conventional tokamaks. It improves confidence in tokamak engineering tolerances, which are a significant driver of cost and time constraints on device construction. We add J-TEXT data, new JET data, and create the scaling using only conventional tokamak Ohmic and L-mode experiments. Since H-mode plasmas are more resilient to error field penetration, this scaling predicts what is likely the most dangerous regime of error field penetration for new tokamak designs. These decisions improve confidence in the error field penetration threshold scaling and its application in the construction and design decisions of any future conventional tokamak or fusion pilot plant.
\end{abstract}

\maketitle
\ioptwocol


\section{\label{sec:Background} Error Field Scaling Background}
\subsection{Tokamak Error Fields}
While tokamaks are ideally considered to be axisymmetric, in reality, tilts and shifts in the magnetic field coils, leads and coil windings, the placement of magnetized components, and other various sources create non-axisymmetric error fields \cite{Luxon2003,Ferraro2019ErrorNSTX-U,Piron2020ErrorOperation,Menard2010ProgressPlasmas}. When these fields, either alone or combined with the applied 3D field from the 3D coils of a given tokamak, exceed the error field penetration threshold in a given device, they can lead to disruptions by generating magnetic islands through magnetic reconnection and then locking those islands in the rest frame \cite{LaHaye1992,Fitzpatrick2012NonlinearPlasmas}. Error fields can also cause difficulties for applied 3D field experiments, such as resonant magnetic perturbation edge localized mode suppression \cite{Evans2004,Schaffer2011,Lunia2026}. Experimentally, it has been observed that low-wavelength, particularly n=1 error fields are the primary cause of these effects \cite{Lanctot2017ImpactTokamaks,Buttery2000ErrorJET,Menard2010ProgressPlasmas}.

Since these disruptive locked modes must be avoided to achieve robust tokamak operation, the error field penetration threshold provides constraints on the design and construction of tokamaks, such that the level of error field that would create locked modes at the operating point of a given tokamak is above the level of expected error field produced from the device. The resulting engineering tolerances for the tokamak place constraints on the cost and build time of new devices. These considerations are especially important for devices on the scale of ITER \cite{pharr_error_2024,amoskov_optimization_2015} and SPARC \cite{logan_sparc_2026,Sweeney2020MHDTokamak} due to the greater amount of stored energy and thus more dangerous disruptions \cite{Sweeney2020MHDTokamak,Strait2018ProgressITER, lehnen_disruptions_2015}. 

The history of DIII-D showcases how critical these issues are. Since DIII-D was constructed prior to our current understanding of non-axisymmetric error fields, it has a comparatively large intrinsic error field that causes frequent locked modes and disruption without the intervention of 3D error field correction coils \cite{Luxon2002ATokamak}. The primary error field sources for DIII-D, coil tilts and shifts, as well as the buswork (see Reference \cite{Luxon2003}), can be avoided or compensated for through more careful design and construction on new tokamaks. This is evidenced by the diligent tolerancing efforts, addition of correction coils, and correspondingly relatively small error fields reported in EAST \cite{Yang2018}, JET \cite{FishpoolHaynes1994_FieldErrorInstabilitiesJET}, C-MOD \cite{Vieira2003_CModNonAxisymmetricCoils}, J-TEXT \cite{Rao2013_MeasurementIntrinsicErrorFieldJTEXT},  and KSTAR \cite{in_extremely_2015}.

\subsection{\label{old_scale}Existing Database and Scaling}
The leading standard in empirical error field penetration threshold prediction is producing scalings from database studies of tokamak shots with locked modes \cite{Buttery1999,Park20172017ITER,logan_empirical_2020,logan_robustness_2020}. Note that although scalings for the n=2 threshold exist \cite{logan_empirical_2020}, this paper focuses only on the n=1 error field penetration threshold, since it remains the most important harmonic with respect to disruption risk \cite{Lanctot2017ImpactTokamaks}, and there is more data available. These scalings relate the core dominant mode overlap, $\delta$ \cite{Park2011ErrorHandedness}, to a set of plasma parameters. 

Resonant fields at rational surfaces drive the onset of island opening \cite{LaHaye2006NeoclassicalControl,Park2007ControlTokamaks,Park2007ComputationEquilibria}, and the core dominant mode overlap quantifies how strongly the applied field in a tokamak overlaps with the field that would maximally drive tearing, while accounting for the plasma response. For a given set of plasma conditions and 3D coil configuration, we compute $\delta$ using the General Perturbed Equilibrium Code (GPEC) \cite{Park2007ComputationEquilibria,Park2009ShieldingPlasmas,Park2009ImportanceTokamaks}.

The overlap metric was originally defined in \cite{Park2011ErrorHandedness,park_corrigendum_2012} as a fractional quantity, the percent overlap $\mathcal{C}$, ranging from 0 to 1: $\mathcal{C}=1$ when the external field normal to the control surface exactly matches the dominant external field normal to that surface, and $\mathcal{C}=0$ when the two fields are orthogonal. Later work \cite{logan_empirical_2020,logan_robustness_2020,bandyopadhyay_mhd_2025}, including the present study, instead adopts the core dominant mode overlap, $\delta$, which simply expresses the overlap, $\mathcal{C}$, with surfaces within $\psi_n = 0.9$ and as an absolute quantity rather than a percentage (where $\psi_n$ is the normalized poloidal flux). 

The overlap metric follows from the insight that a physically meaningful resonant metric must be independent of the choice of coordinate system \cite{Pharr2026CoordinateInvariant}. Reference \cite{Park2008SpectralCoordinates} shows that the pitch-resonant Fourier component of the surface-normal magnetic field is coordinate-system invariant at its corresponding rational surfaces $q=m/n$ when written with an area-normalization as

\begin{equation}
    \bar{b}_{m n} \equiv \frac{1}{\oiint d\text{A}} \oiint d \theta \, d \phi \,\delta \mathbf{B} \cdot \hat{n} \mathcal{J}|\nabla \psi| e^{i(n \phi-m \theta)}
\end{equation}

\noindent where $\mathcal{J}\left|\nabla \psi\right| d\theta\,d\phi$ is the surface area element, $\hat n$ is the flux-surface-normal unit vector, and $\delta \mathbf{B}$ is the perturbed total magnetic field. This can, in turn, be related by a matrix equation to a given root-area-normalized external field, $\tilde{b}^x_{mn}$,

\begin{equation}
    \tilde{b}_{m n}^x=\frac{1}{\sqrt{\oiint d\text{A}}} \oiint d \theta \,d \phi \,\delta \mathbf{B}^x \cdot \hat{n} \sqrt{\mathcal{J}|\nabla \psi|} e^{i(n \phi-m \theta)}
\end{equation}

\noindent Here $\delta \mathbf{B}^x$ is the perturbed vacuum field, and we are now integrating over an external control surface. Considering a single toroidal component ($n=1$ for this paper), the matrix equation linking the two quantities is

\begin{equation}
    \bar{b}_i^r=C_{i m} \tilde{b}_m^x.
\end{equation}

\noindent where $C$ is a coupling matrix, whose long dimension is the number of retained poloidal modes and whose short dimension is the number of rational surfaces contained within $\psi_n = 0.9$. Taking the singular value decomposition (SVD) of C,

\begin{equation}
    C_{i m}=U_{i p} \Sigma_{p q} V_{q m}^*
\end{equation}

\noindent and noting that the first singular value is typically much larger than the rest, we approximate the full coupling matrix by its first right singular vector, $\mathbf{v}_1^\dagger$ \cite{Park2008ErrorITER,logan_empirical_2020}. The dimensionless dominant mode overlap metric, $\delta$, then follows from the dot product of $\mathbf{v}_1^\dagger$ with $\mathbf{\tilde{b}}^x$, normalized by the toroidal magnetic field, $B_T$ (in T):

\begin{equation}
    \delta = \frac{|\mathbf{v}_1^\dagger \cdot \mathbf{\tilde{b}}^x|}{B_T}
    \label{eq:delta_dot_prod}
\end{equation}

\noindent A detailed treatment of the metric and proof it is coordinate system invariant is given in \cite{Pharr2026CoordinateInvariant}. Since we use this metric across a wide range of device sizes and field strengths and use the metric to project to a range of devices that can have sizes and field strengths outside of existing parameter ranges, it is essential that this metric be dimensionless.

The dominant mode overlap, unlike other metrics, such as the three-mode metric \cite{Scoville2003Multi-modeTokamak},  is coordinate system independent, matches well with experimental evidence, and accounts for the plasma response. Since the metric is device-agnostic, it can be used for multi-machine scaling laws that provide predictive capabilities for new tokamak designs. It has been used widely on many devices to understand error field penetration in tokamak plasmas \cite{Boozer2015, paz-soldan_importance_2014, logan_empirical_2020, logan_robustness_2020, Yang2021ParametricPlasmas}.

To produce a scaling law, the dominant mode overlap is fit using a log-linear regression power law scaling from the error field penetration threshold database. The database used in References \cite{logan_robustness_2020} and \cite{bandyopadhyay_mhd_2025} consists of a combination of steady-state H-mode, L-mode, and Ohmic plasmas across DIII-D, KSTAR, EAST, C-MOD, COMPASS, JET, and NSTX. (Note that this work excludes all H-mode cases, as well as NSTX and COMPASS discharges, while incorporating additional data, including new J-TEXT discharges, as described in Section \ref{sec:updates}.). The Ohmic plasmas are primarily in linear Ohmic confinement (LOC) as opposed to saturated Ohmic confinement (SOC). A piecewise scaling or pure SOC scaling is left for future work. As shown in \cite{Yang2021ParametricPlasmas}, the error field density dependence changes dramatically depending on the confinement regime. Since it includes NSTX, it spans both traditional and spherical tokamaks (although it is primarily traditional tokamaks). It also incorporates both high field side 3D coils and low field side 3D coils, which are thought to differ in their relationship to tearing modes. 

The incorporation of intrinsic error field effects varies significantly across devices in the earlier database. Comprehensive models for n=1 intrinsic error sources have been developed for NSTX and DIII-D. These models were fully utilized for all NSTX cases \cite{Park2007ControlTokamaks,Menard2010ProgressPlasmas}, whereas for DIII-D only a limited number of well-reconstructed equilibrium cases were used to approximate the intrinsic error fields \cite{Park2011ErrorHandedness}. For all other devices, the intrinsic error fields are either ignored or approximated by subtracting the empirically optimal correction coil currents. A list of device error fields can be found in Table 1 of Reference \cite{in_extremely_2015}. The database was compiled from different sources across several years, resulting in a variety of GPEC versions used to compute $\delta$. It also comprises calculations using different types of Jacobians, which in principle should not alter the coordinate-invariant resonant fields \cite{Park2008SpectralCoordinates} but could introduce numerical truncation errors that vary inconsistently across the database. These variations can contribute as much as a 30\% difference in $\delta$ for any individual shot.

For the most part, there is not a direct comparison between empirical database scalings that include any Ohmic and L-mode data and theoretical scalings, see References \cite{Fitzpatrick2012NonlinearPlasmas,Cole2006,Cole2007,Cole2008}. This is because the majority of the database lacks neutral beam data (necessary for charge exchange measurements of rotation) and other diagnostics necessary for such scalings. We therefore do not have the requisite data needed to compute one-to-one scaling comparisons.

In the following work, $n_e$ is the electron density, $I_p$ is the plasma current, $R_0$ is the major radius, $a$ is the minor radius, $A$ is the plasma control surface area, $B_T$ is the toroidal magnetic field, $\beta_n$ is the normalized plasma pressure, $l_i$ is the internal inductance, $\kappa$ is the elongation, $\delta_T$ is the triangularity, $\zeta$ is the squareness, and $q_{95}$ is the safety factor at 95 percent of the poloidal flux.

This paper seeks to maximize the fit quality, expand the parameter space, and improve predictive confidence in the error field penetration threshold scaling. Across the previous log-linear regression power law scalings \cite{logan_robustness_2020,logan_empirical_2020,bandyopadhyay_mhd_2025} $R^2$, the coefficient of determination, has values as low as 0.34 and uncertainty on the ITER penetration threshold as high as 48 percent. By improving the scaling through a standardized database, new fitting techniques, and an expanded parameter space, we seek to provide a more reliable prediction of when error fields will produce deleterious effects in future conventional tokamaks.

The remainder of the paper is set up as follows. Section 2 describes changes made to the database and the error field scaling, Section 3 introduces the new scaling and how it compares to others, and Section 4 discusses the quantification of uncertainty. Section 5 looks at scalings made up of reduced sets of devices, Section 6 uses Monte-Carlo methods to project to ITER, and Section 7 concludes with a discussion of possible improvements and outlook.

\section{\label{sec:updates} Updates to the Error Field Penetration Threshold Database}
\subsection{Regime and Simulation Selections}
Performing analysis and making predictions using databases inevitably leads to the inclusion of various known, unknown, systematic, and random uncertainty. These can arise from device-specific data or operational idiosyncrasies and variations in the modeling and computational analysis schemes. Also, while the scaling is meant to encompass as large a parameter space as possible, it is better to avoid drastic changes between physics operating regimes. 

In that vein, this paper focuses on scaling only with L-mode and Ohmic cases from conventional tokamak experiments.  Therefore, all previous H-mode discharges are excluded from this work. Since H-mode shots have higher densities, they are typically much safer with respect to the penetration threshold \cite{logan_sparc_2026}. Of the Ohmic confinement shots, we exclude SOC shots from density scans to avoid confounding the density dependence by spanning the LOC to SOC threshold \cite{Yang2021ParametricPlasmas}. This means that the scaling is better suited for applications where LOC is expected, rather than SOC. Since all tokamaks are able to operate in some amount of LOC, particularly in the plasma current ramp-up, this scaling can be applied to LOC operation in any future conventional tokamak. The scalings now only include conventional, non-spherical tokamaks and only data from low-field-side error field experiments. Although this means that the scaling is less suited than previous scalings for very low inverse aspect ratio tokamaks, it is better suited than previous scalings for conventional tokamaks. The devices in the scaling span inverse aspect ratios of 0.24 to 0.36, encompassing the ITER and SPARC inverse aspect ratios of 0.32 and 0.31, respectively \cite{Creely2020OverviewTokamak}. Since we are prioritizing confident extrapolation to ITER and SPARC, the more similar the data is to future devices being planned and built, the higher confidence we can have in the predictive capacity of the scaling.

DIII-D's intrinsic error field is known to be one of the largest of any currently operating tokamak and this work incorporates a direct model of the known error field sources into the overlap calculation. The previous database incorporated intrinsic error field contributions for DIII-D using SURFMN \cite{Schaffer2008} applied to the known coil geometry \cite{Luxon2003}. However, rather than computing individual equilibrium reconstructions and intrinsic error field calculations for each shot, shots with similar plasma conditions were grouped together, with a single representative equilibrium and intrinsic error field applied to all shots within each group. The current database improves upon this approximation by computing dedicated equilibrium reconstructions and SURFMN-based intrinsic error field calculations for each DIII-D shot individually, ensuring that the penetration threshold is not artificially biased by errors introduced through the grouping approximation. This refinement does, however, retain the added uncertainty inherited from the modeling of the coil tilts and shifts. The remaining devices assume a negligible error field or error field corrected to negligible levels.

\subsection{New Data}
\begin{figure*}[h]
\centering{}\includegraphics[width=1\linewidth]{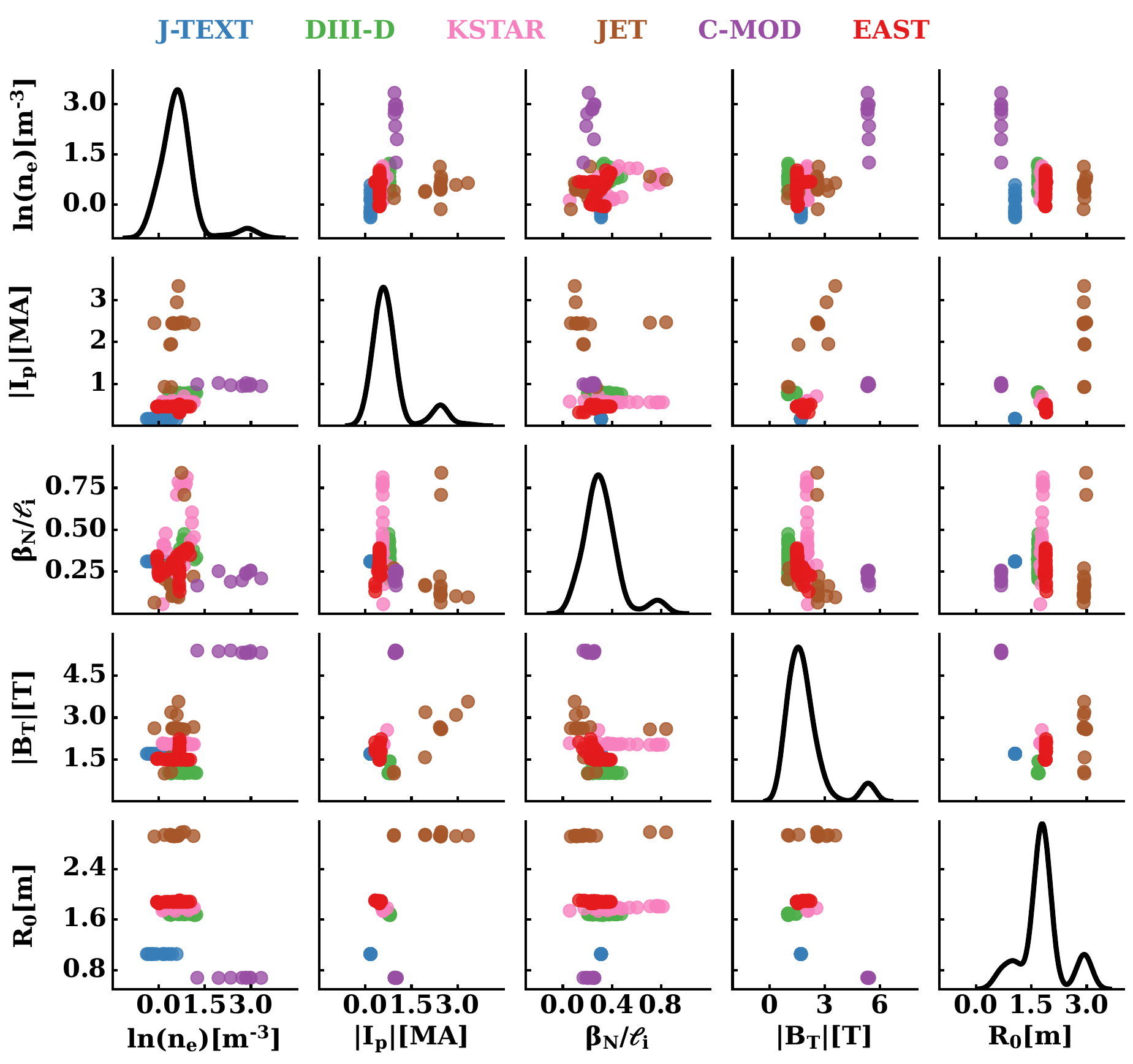}\protect\caption{Parameter space plot demonstrating additions of J-TEXT and JET and absence of spherical tori. Plots along the diagonal are the distributions of single parameters, to highlight the skewness of the data. Section \ref{sec:new_param_combo} provides scalings with and without weighting based on data density distributions. The database contains 41 EAST shots, 23 DIII-D shots, 22 KSTAR shots, 19 JET shots, 13 J-TEXT shots, and 9 C-MOD shots.  \label{fig:param_space}}
\end{figure*}

The scope of the database parameter space is a large lever on the confidence in projections to new tokamaks or scenarios. Ideally, predictions using the scaling would fall within the database parameter space. However, given the impossibility of this, particularly for the major radius of ITER, we would like to have as great a variety as possible in parameter space to increase the confidence in the scaling projecting accurately to new devices. Figure \ref{fig:param_space} shows the distribution of key scaling parameters across the database for each device. The diagonal line plots show the distribution of individual parameters, with single peaked, broad distributions being more desirable for avoiding sampling biases in an empirical scaling.

This work includes J-TEXT data for the first time and considers its effect on the projection to ITER. Its major radius size is especially important since the focus on conventional tokamaks removes NSTX which is of a similar size. J-TEXT also contributes data at low plasma current and across a strong range of low density, providing new minimum values for both parameters. Unlike the other database devices, J-TEXT does not have routine equilibrium reconstructions, so a representative equilibrium was reconstructed using the code TokaMaker \cite{hansen_tokamaker_2024}. This method allows us to convert 3D coil currents to dominant mode overlap using GPEC. J-TEXT has strong density scaling data at an important major radius \cite{Mao2022StudyJ-TEXT,Wang2014StudyJ-TEXT,Huang2020DisentanglementTokamak}, as seen in the blue points in Figure \ref{fig:param_space}. Other than J-TEXT, all tokamak discharges use magnetic equilibria reconstructions. The parameters in the scaling are derived from their equilibria with the exception of density. 

We also introduce an expanded shot range for JET, from only 4 shots to 19. This is important since JET is an outlier in large major radius, which is a critical parameter for ITER and fusion pilot plant, FPP, extrapolation. These shots are from an error field experiment using the JET saddle coils to simulate error fields and determine penetration thresholds with scans of density and toroidal field \cite{Buttery2000ErrorJET}. They are included for the first time now, thanks to a conversion process that translates old magnetic data into a form that is compatible with EFIT \cite{Lao1985ReconstructionTokamaks} reconstruction. As seen in the brown points in Figure \ref{fig:param_space}, this JET data is well above all other major radius points, and also spans a wide range of plasma current, well above most other devices. The JET data also spans the widest range of $\beta_n/l_i$, normalized plasma pressure, of any device.

\subsection{Parameter Selection Expansion}
Previously, the error field scaling was fit using density, toroidal field, major radius, and $\beta_n$ over internal inductance. This choice was guided by linear tearing threshold theories \cite{Cole2006,Fitzpatrick1991} and the strongest trends observed in individual experiments \cite{Wang2014StudyJ-TEXT,Mao2022StudyJ-TEXT,Buttery2000ErrorJET,Park2011ErrorHandedness,Wang2020ToroidalEAST,Wolfe2005NonaxisymmetricC-Mod,in_extremely_2015}. Since the dominant mode overlap is calculated using the first singular vector and dropping the singular value, we lose some information contained in the singular value such as the dimensionality. This improves numerical robustness, but can leave the scaling missing important shaping and plasma amplification information, necessitating the addition of terms in other scalings like $\beta_n$\cite{Park20172017ITER,logan_robustness_2020,logan_empirical_2020}. We can open up the parameter space to better account for the physics by including new shaping and geometry terms, explicit $q_{95}$ dependence, and plasma current dependence. The full list of available parameters can be found in Table \ref{tab:param_table}.

\begin{table*}[t]
\centering
\renewcommand{\arraystretch}{1.15}
\setlength{\tabcolsep}{6pt}
\resizebox{\textwidth}{!}{%
\begin{tabular}{lcccccccccccc}
\hline
& $n_e$ [$10^{19}$ m$^{-3}$] 
& $I_p$ [MA] 
& $R_0$ [m] 
& $a$ [m] 
& $A$ [m$^2$]
& $B_{T}$ [T]
& $\beta_n$
& $l_i$
& $\kappa$
& $\delta_T$
& $\zeta$
& $q_{95}$ \\
\hline
2010, 2017 \cite{bandyopadhyay_mhd_2025}
& $\checkmark$ 
&  
& $\checkmark$ 
&  
&  
& $\checkmark$
& $\checkmark$
& 
&  
&
&
&   \\
2020 \cite{logan_robustness_2020}
& $\checkmark$ 
&  
& $\checkmark$ 
&  
&  
& $\checkmark$
& $\checkmark$
& $\checkmark$
&  
&
&
&   \\
Present Work
& $\checkmark$ 
& $\checkmark$ 
& $\checkmark$ 
& $\checkmark$ 
& $\checkmark$
& $\checkmark$
& $\checkmark$
& $\checkmark$
& $\checkmark$
& $\checkmark$
& $\checkmark$
& $\checkmark$ \\
Maximum Value
& $28$
& $3.3$
& $3$
& $1$
& $143$
& $5.4$
& $1$
& $1.8$
& $1.7$
& $0.6$
& $0$
& $6.5$ \\
Minimum Value
& $0.7$
& $0.2$
& $0.7$
& $0.2$
& $6.9$
& $1$
& $0.1$
& $1$
& $1$
& $0$
& $-0.1$
& $2.7$ \\
\hline
\end{tabular}%
}
\caption{Plasma parameters used for scaling in References \cite{logan_robustness_2020,bandyopadhyay_mhd_2025} versus those investigated for fitting in this work. The maximum and minimum values of each parameter are included to identify for future design points whether a chosen parameter is within the bounds of the existing data.}
\label{tab:param_table}
\end{table*}
\subsection{Distribution of Observed Thresholds}
\label{sec:distribution}
\begin{figure}[h]
\includegraphics[width=1\linewidth]{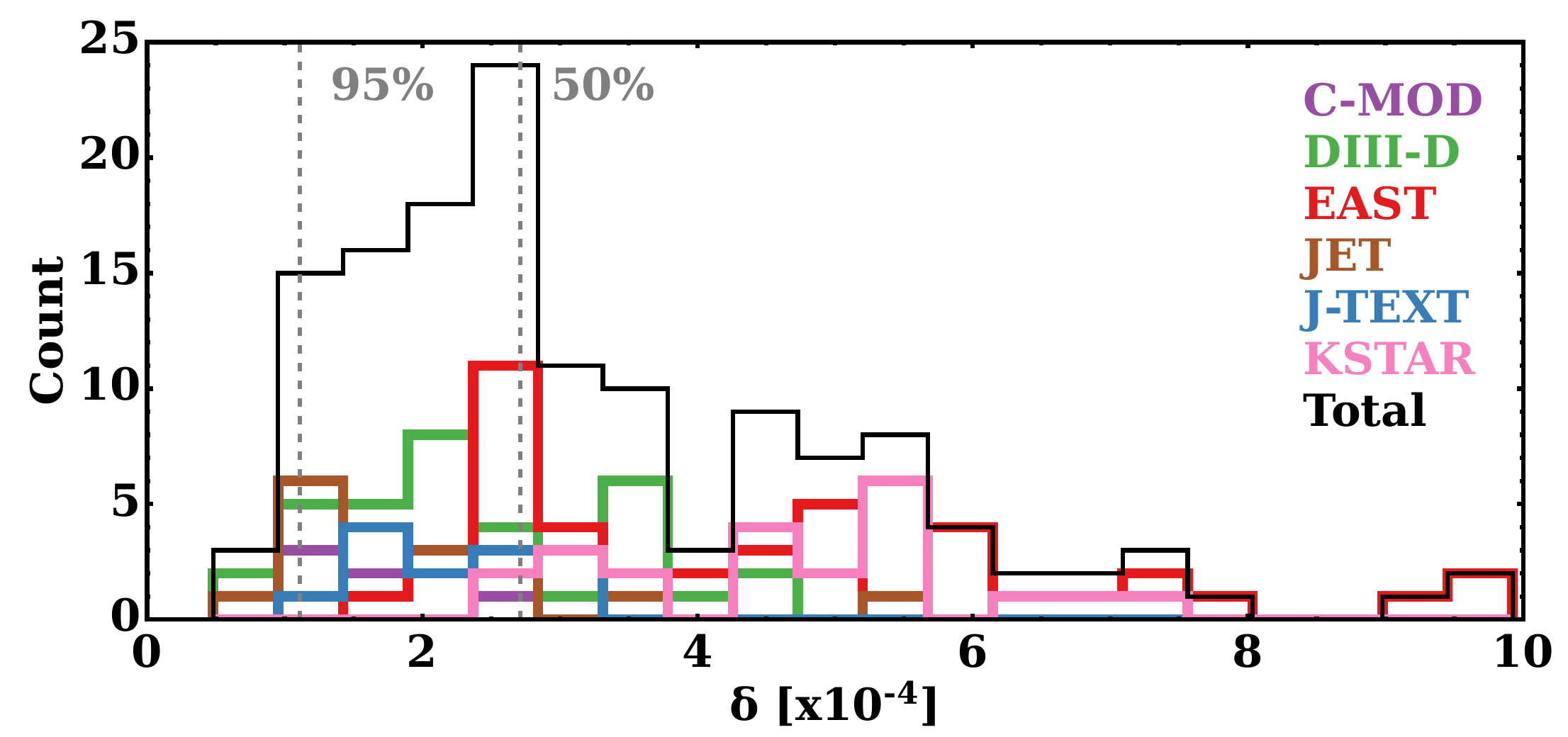}\protect\caption{Distribution of the error field penetration threshold scaling database. The y-axis is the number of discharges in each bin and the x-axis the core dominant mode overlap metric at which error field penetration was observed for each point. \label{fig:ef_dist}}
\end{figure}

The distribution of the dominant mode overlap at which each discharge in the database underwent error field penetration can be seen in Figure \ref{fig:ef_dist}. Although the purpose of this paper is to provide an empirical error field penetration threshold scaling, a simpler method is to consider the distribution of observed error field penetration thresholds and design future devices to be below some percent of the dataset. $99\%$ of the database experienced error field penetration above $\delta = 5.98\times10^{-5}$, $95\%$ of the database experienced error field penetration above $\delta = 1.11\times10^{-4}$, and $50\%$ of the database experienced error field penetration above $\delta = 2.71\times10^{-4}$. As an example, Reference \cite{logan_sparc_2026} shows that although SPARC is designed using the error field penetration threshold scaling, the scaling predicts that the threshold for SPARC's operating points is lower than most observed thresholds.

\section{\label{sec:new_param_combo} New Error Field Penetration Threshold Scaling}
\subsection{Scaling Methodology}
Taking all of the parameters, a forward selection least-squares regression is used to find the best possible log-linear power law. We use least-squares and the log-linear power law form due to their proven success and to maintain continuity with past error field penetration scalings \cite{bandyopadhyay_mhd_2025}. This also connects this work to other successful tokamak scalings such as the ITER $H_{98y2}$ \cite{transport_chapter_1999}. Perhaps most importantly, the theoretical scalings for error field penetration produce power law results as well \cite{Cole2006,Fitzpatrick2012NonlinearPlasmas,Fitzpatrick2003PlasmaTokamaks}. Optimizing the scaling outside of these constraints could prove fruitful for future work, though it is outside the scope of this paper.

First, each parameter is fit alone to determine which produces the fit with the highest $R^2$. That parameter is then selected and refit with each of the remaining parameters to find the best fit with two variables. This carries on until either a set number of parameters is reached, a maximum $R^2$ increase of $10^{-4}$ is not found, or the condition number exceeds $20$. A minimum of two variables is also set.  In Equations \ref{eq:deltau_UQ} and \ref{eq:delta_KDE}, the fit was stopped after selecting 5 variables to keep the condition number below 20. The coefficient of determination, $R^2$, is defined as \cite{GlantzSlinker1990}: 
\begin{equation}
    R^2 = 1-\frac{SS_{\text{res}}}{SS_{\text{tot}}}
\end{equation}
\noindent $SS_{\text{res}}$ is the sum of squares of the residual and $SS_{\text{tot}}$ is the total sum of squares. The closer $R^2$ is to one, the better the fit describes the data. The condition number indicates how far a matrix is from being singular, which is correlated with degree of sensitivity to small perturbations of inputs. Also, in each step, if a parameter is less than 0.01 $R^2$ away from the best, but has less collinearity as indicated by a lower condition number, that parameter is selected instead. 

\begin{figure}[h]
\centering{}\includegraphics[width=1\linewidth]{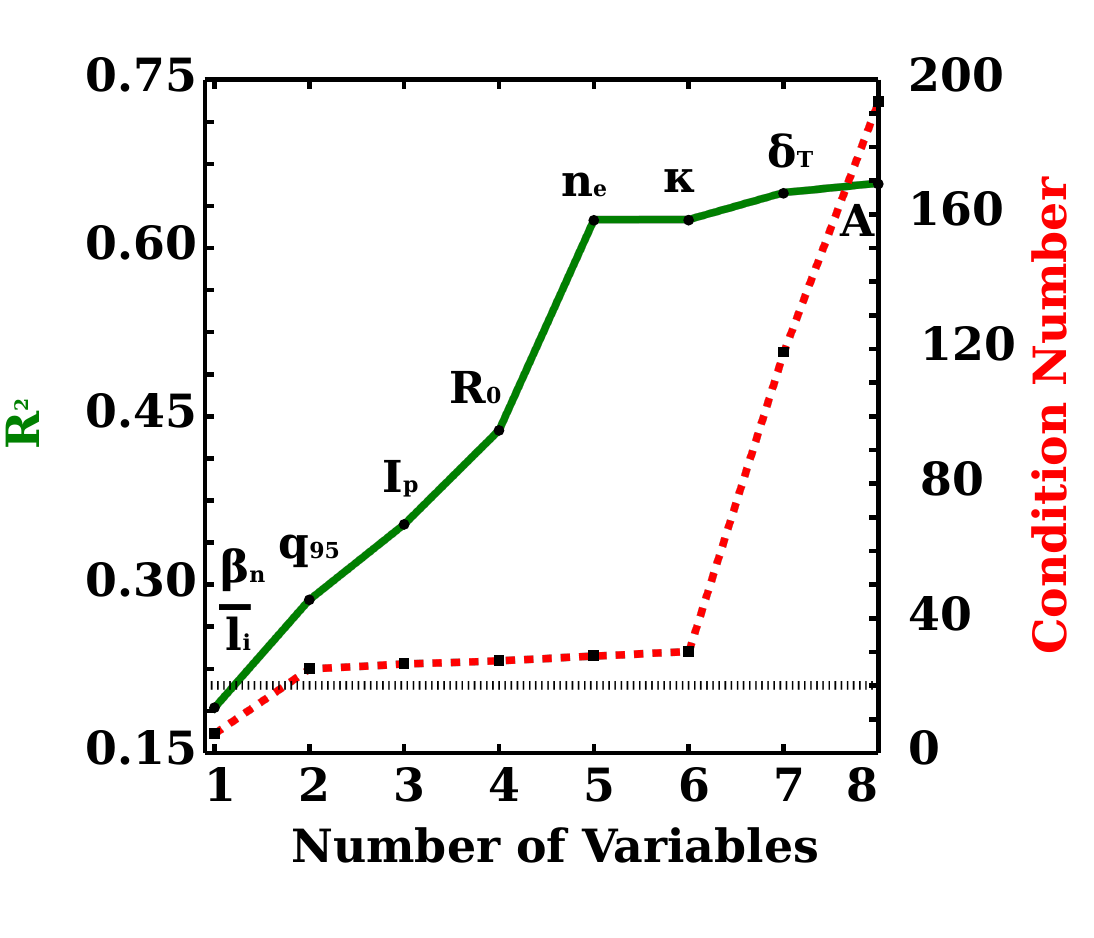}\protect\caption{$R^2$ and condition number versus number of parameters added. Note that this is an illustrative plot showing trade-offs between $R^2$ and condition number. It is not representative of the order of variables implemented, since the actual scalings are limited in variable number due to the limit on condition number. The number of variables is truncated at 8 for visualization as the condition number becomes overwhelmingly large beyond this ($1.8\times10^3$ at 9 variables). \label{fig:fitquality}}
\end{figure}

In Figure \ref{fig:fitquality}, the parameters are fit one by one, recording $R^2$ and condition number at each step. While the figure illustrates trade-offs between $R^2$ and condition number, since the final scalings produced here use a limit of 20 on the condition number, the actual selection of variables in the figure is not identical to that of the scalings presented. Interestingly, the variables are added in order of largest $R^2$ contribution, but despite the large contribution from $R_0$, it only increases the $R^2$ to a much larger degree after the prior three variables are added. With the condition number limited to 20 and minimum variables set to 2, the order changes from $\beta_n/l_i$, $q_{95}$, $I_p$, $R_0$, $n_e$, $\kappa$, $\delta_T$, and A, (as seen in Figure \ref{fig:fitquality}), to $\beta_n/l_i$, $I_p$, $R_0$, $n_e$, and $B_T$ (as reflected in Equations \ref{eq:deltau_UQ} and \ref{eq:delta_KDE}). 

\begin{figure}
    \centering
    \includegraphics[width=1\linewidth]{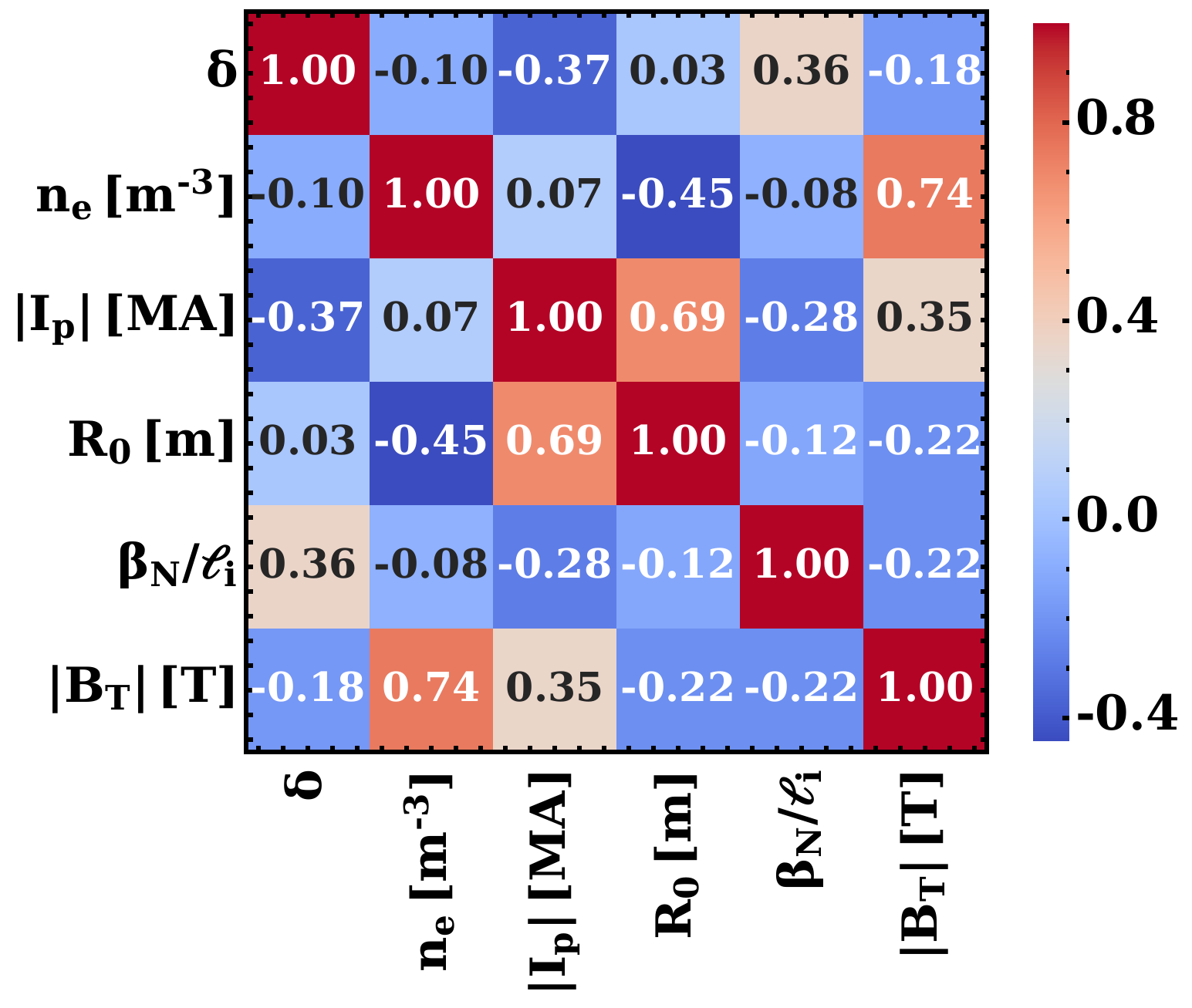}
    \caption{Correlation matrix for a selection of key variables in the reported error field penetration threshold scaling equations.}
    \label{fig:corr_mat}
\end{figure}

\subsection{Parameter Correlation}
To better understand the dependencies between each variable, it can be helpful to examine a correlation matrix for the strength of the correlation of each variable with $\delta$ as well as with each other variable. This is found in Figure 
\ref{fig:corr_mat}, which is a 6 by 6 matrix where a correlation of 1 means that two variables are perfectly correlated, here found only for a variable with itself. Note that both strong positive and strong negative correlations improve fit quality in the regression. We see that the variables with the highest correlation with $\delta$ are the plasma current at -\mbox{0.37} and $\beta_n/l_i$ at \mbox{0.36}. This agrees with their selection as the first two variables into the fit. However, the third highest correlated parameter is not major radius, which was selected third, but rather toroidal field, highlighting the importance of secondary dependencies between variables for scaling impact. As the forward selection continues, these mechanisms continue to play out between each variable in ways that are not immediately apparent looking solely at each parameter's individual correlation with $\delta$.

Another important finding from Figure \ref{fig:corr_mat} is the high degree of correlation, 0.74, between density and toroidal field. This is largely due to the uniquely high operating points of C-MOD in both toroidal field space and density space. Adding in additional C-MOD or other data in the intermediary parameter space should decrease the magnitude of this particular correlation feature. 

\subsection{Changes in Scaling Dependencies}
The unweighted fit found with the outlined decisions and scaling technique has $R^2 = 0.63$ and a condition number of 17.76:

\begin{equation}
\label{eq:deltau_UQ}
\begin{split}
\delta = 10^{-4.31\pm0.09}
\left(\frac{\beta_n}{l_i}\right)^{0.25\pm0.08}
|I_p|^{-0.97\pm0.08}
R_0^{1.88\pm0.16}\\
n_e^{0.77\pm0.08}
|B_T|^{0.19\pm0.09}
\end{split}
\end{equation}

Instead of the outlined ordinary least squares method, we can also consider using a regression weighted using a kernel density estimate (KDE), as explained in \cite{logan_robustness_2020}, to avoid oversampling data-rich parts of the parameter space. This method, with the parameters chosen in Equation \ref{eq:deltau_UQ}, produces a scaling with $R^2 = 0.66$ and a condition number of 17.76:

\begin{equation}
\label{eq:delta_KDE}
\begin{split}
    \delta = 10^{-4.26\pm0.09}\left(\frac{\beta_n}{l_i}\right)^{0.13\pm 0.06}|I_p|^{-1.01\pm0.07}R_0^{1.57\pm 0.15}\\ n_e^{0.56\pm 0.08} |B_T|^{0.30\pm 0.10}  
\end{split}
\end{equation}

\noindent As mentioned before, the past scalings had values of $R^2$ as low as 0.34 and the scalings presented here are robustly as good or better than any previously presented scalings. Note that the uncertainty quantification is discussed in Section \ref{sec:UQ}. We can compare these to ordinary least squares (OLS) and weighted least squares (WLS) Ohmic and L-mode (O,L) n=1 scalings reported in 2020 \cite{logan_robustness_2020}, as well as the OLS O,L scalings from 2010 and 2017 \cite{bandyopadhyay_mhd_2025}:
\begin{table}[h]
\centering
\begin{adjustbox}{width=\columnwidth}
\begin{tabular}{lcccccc}
\hline
Scaling & $n_e$ & $B_T$ & $R_0$ & $\beta_n$ & $I_p$ & $\left(\frac{\beta_n}{l_i}\right)$ \\
\hline
2010 OLS & 1.30 & -2.00 & 0.93 & -0.69 & -- & -- \\
2017 OLS & 1.40 & -1.80 & 0.81 & -0.86 & -- & -- \\
2020 OLS & 0.63 & -0.98 & 0.15 & -- & -- & -0.13 \\
2020 DSOLS & 0.58 & -1.08 & 0.19 & -- & -- & 0.26 \\
2020 WLS & 0.65 & -1.17 & 0.17 & -- & -- & 0.11 \\
Eqn.~\ref{eq:deltau_UQ} OLS & 0.77 & 0.19 & 1.88 & -- & -0.97 & 0.25 \\
Eqn.~\ref{eq:delta_KDE} WLS & 0.56 & 0.30 & 1.57 & -- & -1.01 & 0.13 \\
\hline
\end{tabular}
\end{adjustbox}
\caption{Comparison of the power-law exponents for published empirical error field scaling laws. OLS is ordinary least squares, WLS is weighted least squares, and DSOLS is downsampled ordinary least squares.}
\label{tab:scaling_exponents}
\end{table}

\noindent Here density has units of $10^{19}$ $m^{-3}$, toroidal field is in T, major radius is in m, $I_p$ is in MA, and $\beta_n/l_i$ is dimensionless. The density dependence falls below the 2010 and 2017 scalings, but nearly within the 2020 ranges. The more pronounced differences from the 2010 and 2017 scalings can be understood in terms of both database composition and parameter representation. Those earlier scalings included low-density data points from NSTX and DIII-D cases near the optimal error field correction conditions where the dominant mode overlap is very small and locked modes are observed at substantially low densities. These cases are excluded in the present work as they are susceptible to uncertainties in the intrinsic error field model and may fall into disruption-prone runaway regimes \cite{paz-soldan_non-thermal_2016}. The exclusion of NSTX data, as well as the inclusion of new data from J-TEXT, KSTAR, and EAST, further contribute to the observed differences. Importantly, the apparent discrepancy in the density and toroidal field exponents is substantially reduced when one accounts for the correlation $\beta_n\propto n_e B_T^{-1}R_0$. Substituting this relation into the 2010 and 2017 scalings yields effective exponents more consistent with those reported here. A more complete investigation of these inter-parametric correlations, including analysis across the full database with ST and H-mode cases, is deferred to future work.

The updated scalings, Equations \ref{eq:deltau_UQ} and \ref{eq:delta_KDE}, have density dependence that falls solidly in the center of the spread in data, when compared with the wide array of individual device density scalings. For example, many devices including JET, COMPASS, and DIII-D \cite{Buttery1999} have seen linear density dependencies, $n_e^{1}$. However, J-TEXT has found $n_e^{0.5}$ \cite{Wang2014StudyJ-TEXT}, NSTX has reported density dependence of $n_e^{0.77}$ \cite{Menard2010ProgressPlasmas}, KSTAR has seen $n_e^{0.66}$ \cite{Yang2021ParametricPlasmas}, and EAST has demonstrated $n_e^{0.4}$-$n_e^{0.6}$ \cite{Wang2018DensityEAST,Ye2021DensityTokamak}. It has also been shown that incorporating energy confinement time \cite{Ye2021DensityTokamak}, plasma rotation \cite{Huang2020DisentanglementTokamak}, and the transition from linear Ohmic confinement to Saturated Ohmic confinement \cite{Yang2021ParametricPlasmas} can all change the density scaling exponent on a given device. Given the number of variables at play that can affect the density scaling, it is encouraging that the density dependence is within the previous single-device and database-wide dependencies.

Another important comparison can be made between the empirical error field penetration scaling and theoretical error field penetration scalings. However, there are a wide variety of theoretical density exponents that range from $n_e^{0.25}$ all the way to $n_e^{2}$ \cite{Fitzpatrick2012NonlinearPlasmas,Cole2007,Cole2008}. These variations arise from a combination of regime choice, confinement scaling, and the extent of physics included. The majority of theoretical scalings are not directly comparable to the presented empirical scalings, as mentioned above, due to the lack of temperature and rotation data in the database. However, some theoretical scalings do report engineering parameter versions of their scalings. For example, Reference 
\cite{Fitzpatrick2012NonlinearPlasmas} reports a scaling of:
\begin{equation}
    \left(\frac{b_r}{B_T}\right)_{\mathrm{crit}} \sim n_e \, B_T^{-1.8} \, R_0^{-0.25}
\end{equation}
Yet, this requires specific regime choices that the empirical scaling is not limited to, and so even these are ill-posed for comparison to the error field penetration threshold scaling presented here. Codes such as TM1 \cite{Hu2020NonlinearThreshold} and Slayer \cite{park_parametric_2022} have also been used to produce scaling laws, but not for the n=1 engineering parameter scaling. 

The $R_0$ dependence increases significantly from as low as 0.15 to as high as 1.88, likely due to the exclusion of COMPASS and NSTX data (both of which had higher $\delta$ values but at lower $R_0$), as well as the codependency of $I_p$ and $R_0$ present in the database. The $\beta_n/l_i$ dependence is within the bounds of the 2020 exponents, but much weaker and of opposite sign as compared with the 2010 and 2017 scalings.

Note that the downsampled and weighted Ohmic and L-mode scalings in Reference \cite{logan_robustness_2020} report comparable positive exponents for $\beta_n/l_i$. There is an additional term, $I_p$, that predicts that a lower plasma current is more resilient to error fields. The $B_T$ scaling changes from a strong negative dependence, between -1 and -2, to a small positive dependence, either 0.19 or 0.30 depending on the regression. This is largely due to the new inclusion of $I_p$, as well as the strong influence of NSTX on the 2020 toroidal field scaling. NSTX had notably high values of $\delta$ but possessed a low toroidal field strength, leaving EAST and KSTAR high $\delta$, high $B_T$ shots to change the dependence. Although it is generally assumed that for fixed $q_{95}$ the ratio of $B_T/I_p$ is constant, $q_{95}$ is not held constant across the database. An approximate equation for elongated configurations, with inverse aspect ratio ($\epsilon = a/R_0$), is \cite{editors_chapter_1999}: 
\begin{equation}
    q_{95}=\frac{5a^2B_T}{R_0I_{MA}}\frac{1+\kappa^2(1+2\delta_T^2-1.2\delta_T^3)}{2}\frac{(1.17-0.65\epsilon)}{(1-\epsilon^2)^2}
\end{equation}
From this equation, it is clear that the introduction of $I_p$ does play a role in changing $B_T$; however, the relationship is more complex since there is also a dependence on $R_0$. Also note that the $q_{95}$ in the database is from EFIT, rather than an analytic equation.

\begin{table*}
\centering
\caption{Comparison of scaling exponents for the ITER $H_{98y2}$ confinement scaling, a similar fit for $\delta$, and the $\delta$ fit from Equations \ref{eq:deltau_UQ} and \ref{eq:delta_KDE}. P is the heating power and M is the isotope mass.}
\label{tab:iter_deltau_compare}
\resizebox{\textwidth}{!}{
\begin{tabular}{l|cccccccccc}
\hline
Scaling
& $C$
& $I_p$
& $B_T$
& $n_e$
& $R_0$
& $P$
& $M$
& $\epsilon$
& $\kappa$
& $\beta_n/\ell_i$ \\
\hline
\noalign{\vskip 2.5pt}
ITER $H_{98y2}$
& $-1.25$
& $0.93$
& $0.15$
& $0.41$
& $1.97$
& $-0.69$
& $0.19$
& $0.58$
& $0.78$
&  \\
\hline
\noalign{\vskip 2.5pt}
$\delta$  ($ H_{98y2}$-like) 
& $-4.47$
& $-1.22$
& $0.26$
& $0.81$
& $2.01$
&
&
& $-0.33$
& $0.47$
&  \\
\hline
\noalign{\vskip 2.5pt}
$\delta$ OLS (Eqn. \ref{eq:deltau_UQ})
& $-4.31$
& $-0.97$
& $0.19$
& $0.77$
& $1.88$
&
&
& 
& 
& $0.25$ \\
\hline
\noalign{\vskip 2.5pt}
$\delta$ WLS (Eqn. \ref{eq:delta_KDE})
& $-4.26$
& $-1.01$
& $0.30$
& $0.56$
& $1.57$
&
&
& 
& 
& $0.13$ \\
\end{tabular}}
\end{table*}

If we do a thought exercise of holding $q_{95}$ constant, we recover a strong negative toroidal field dependence more aligned with previous scalings. For example, using Equation \ref{eq:deltau_UQ}, since $\delta \propto B_T^   {0.19}I_p^{-0.97}$ and $q_{95} \propto B_T/I_p$, we find that $\delta \propto B_T^{-0.78} q_{95}^{0.97}$. 

\begin{figure}[h]
\centering{}\includegraphics[width=1\linewidth]{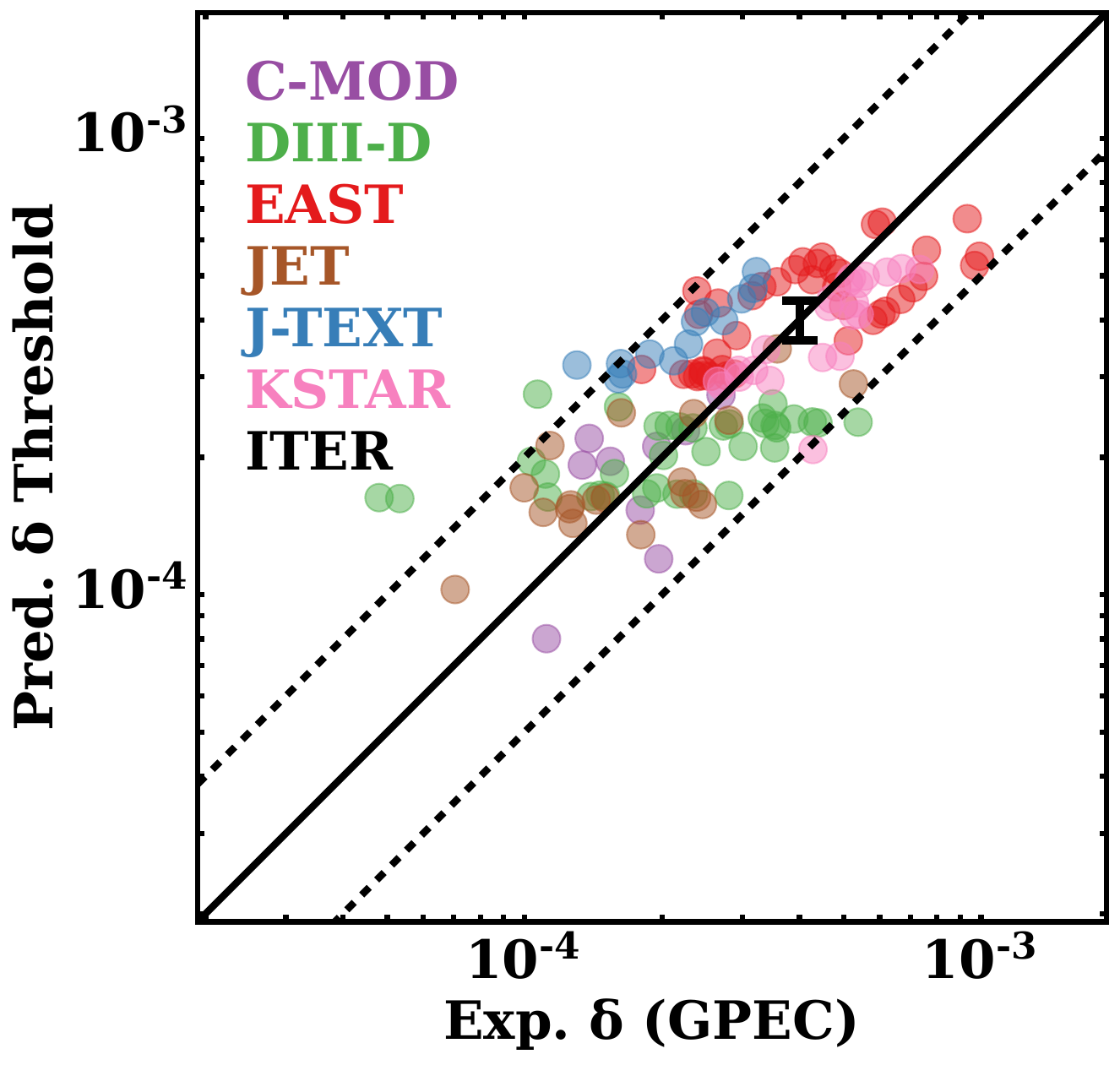}\protect\caption{Best full-device fit using WLS (Equation \ref{eq:delta_KDE}) for the dominant mode overlap ($\delta$)  \label{fig:full_fit}. The perfect correlation (solid) line is surrounded by factor-of-2 lines (dashed) for reference. This fit has an $R^2$ of 0.66, compared to previous scalings with $R^2$ as low as 0.34.}
\end{figure}

Across the parameter space, the exponents change due to a combination of shifts in the datasets and physics dependencies. The new fit containing the parameters from Equation \ref{eq:delta_KDE} is found in Figure \ref{fig:full_fit}. ITER is represented by a vertical error bar, corresponding to the prediction uncertainty, as outlined in Section \ref{sec:UQ}, with parameters defined in Section \ref{sec:ITER_proj}. The linear trend across all devices, as well as the broad scale range of $\delta$ values, demonstrates a suitable quality of fit for predictive purposes. The ITER prediction falling well within the bounds of the existing data and predictions is further reassurance that the ITER prediction is well defined.   

\subsection{Comparison with $H_{98y2}$ Confinement Scaling}
It is also interesting to compare the scaling to the ITER $H_{98y2}$ \cite{transport_chapter_1999} confinement scaling to see broadly whether parameters have similar positive or negative impacts on overall tokamak performance. We consider what will best maximize confinement time and maximize the error field penetration threshold. This comparison is made in Table \ref{tab:iter_deltau_compare}.
\noindent Note that the $\delta \approx$ $H_{98y2}$ fit has $R^2=0.60$, but suffers from some amount of collinearity, as the condition number is 56.8. Density, toroidal field, major radius, and elongation can all be increased in both to improve the metrics (note that the toroidal field in the definition of $\delta$, in Equation \ref{eq:delta_dot_prod}, only increases the positive dependence of toroidal field in the scaling). However, there are conflicting scalings for current and aspect ratio. The main takeaway here is that higher-field, larger, and more dense tokamaks can be generally expected to be more resilient to error fields and have longer thermal energy confinement times. Also, the trends in each parameter are consistent across the OLS, WLS, and $H_{98y2}$-like fits for $\delta$, which is reassuring for the robustness of the scaling.

\subsection{Comparison of $\delta$ with Alternative Error Field Metrics }
\begin{figure*}[tb]
    \centering
    \begin{subfigure}[b]{0.3\linewidth}
        \includegraphics[width=\linewidth]{full_scaling_plot.pdf}
        \caption{Database scaling using the dominant mode overlap (a non-local metric) as defined in Equation \ref{eq:delta_dot_prod}.}
        \label{fig:comp_deltau}
    \end{subfigure}
    \hfill 
    \begin{subfigure}[b]{0.3\linewidth}
        \includegraphics[width=\linewidth]{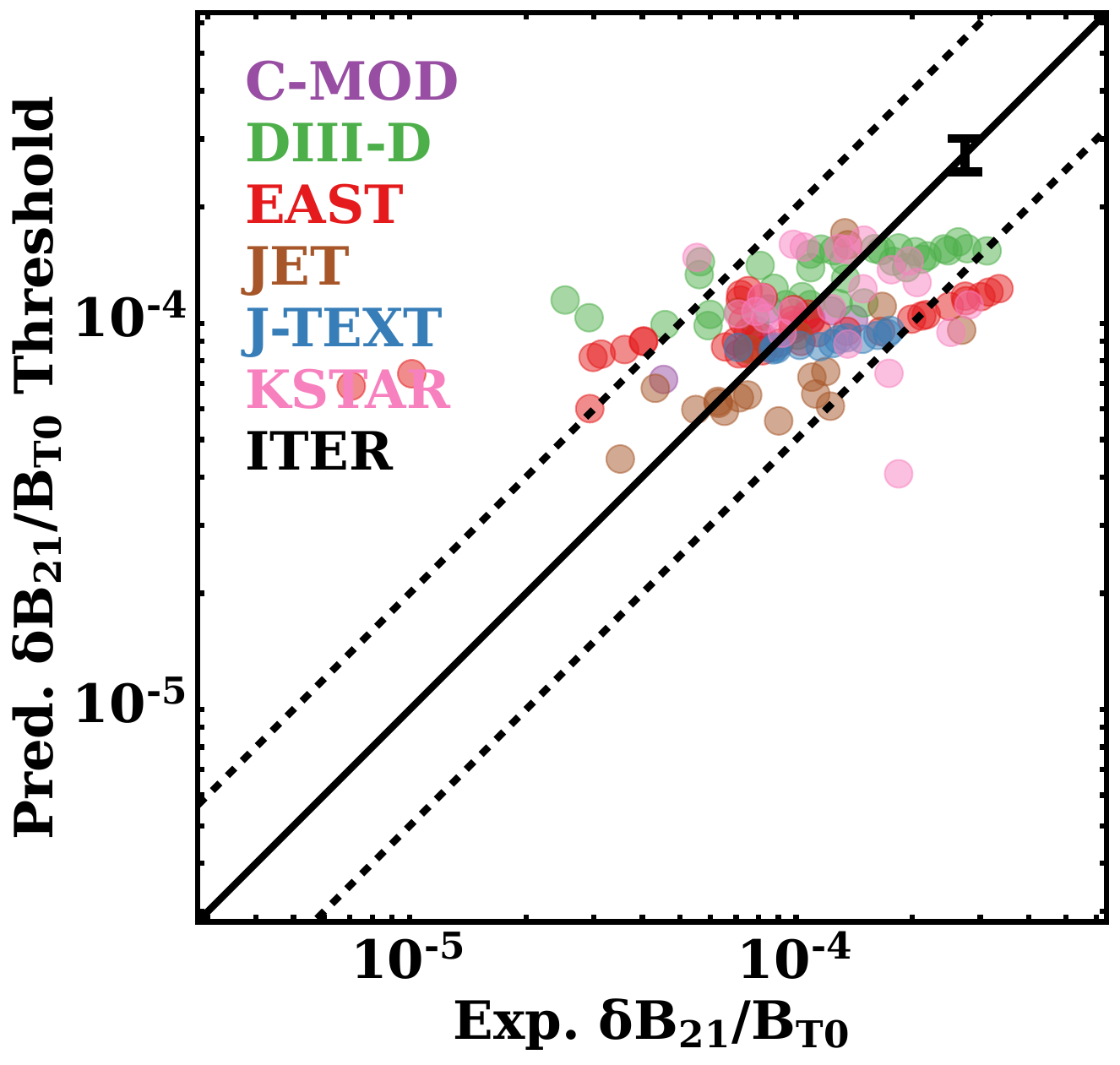}
        \caption{Database scaling using the total resonant flux normalized by the surface area and toroidal field.}
        \label{fig:comp_b21}
    \end{subfigure}
    \hfill 
    \begin{subfigure}[b]{0.312\linewidth}
        \includegraphics[width=\linewidth]{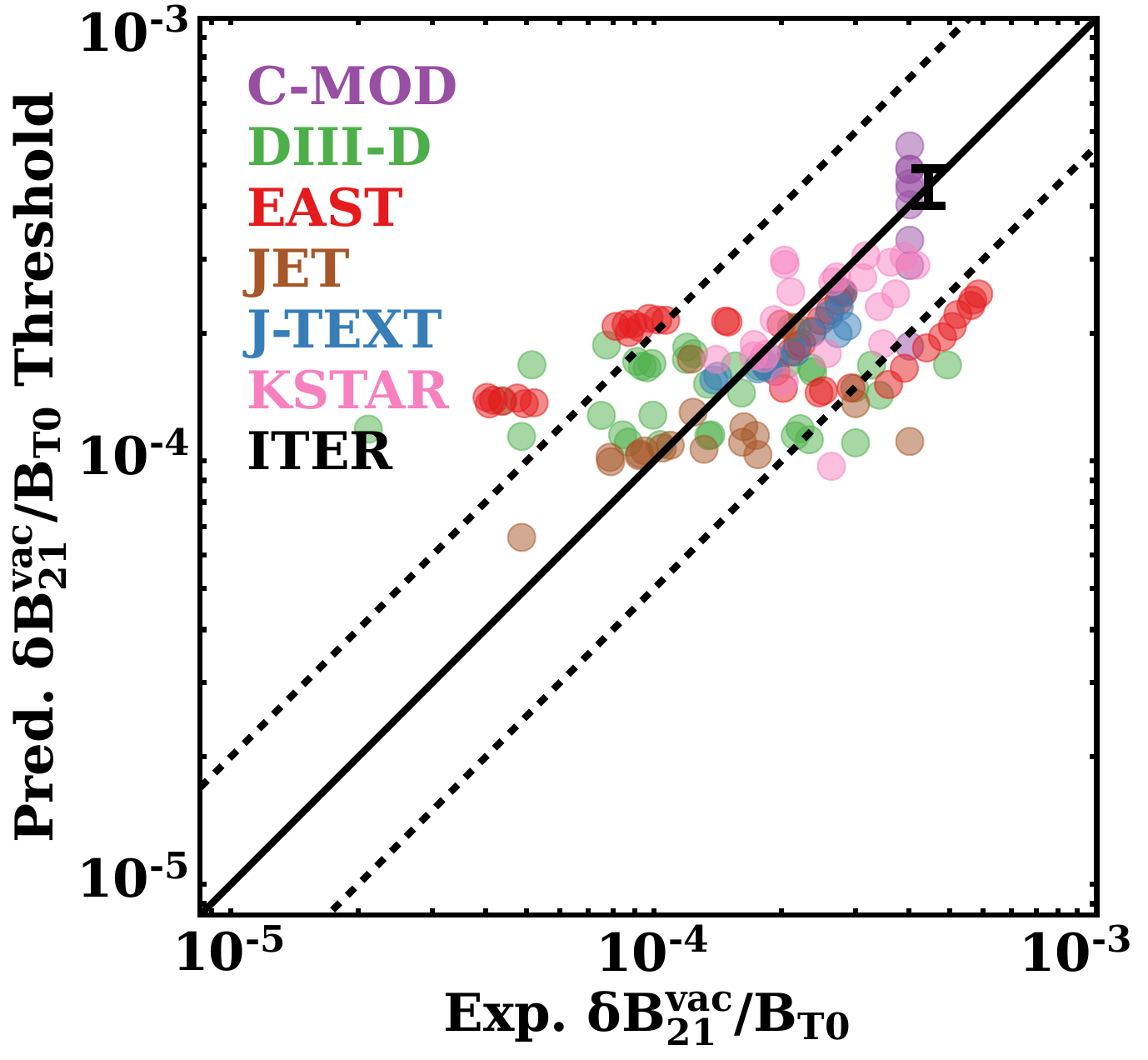}
        \caption{Database scaling using the vacuum resonant flux normalized by the surface area and toroidal field.}
        \label{fig:comp_bv21}
    \end{subfigure}
    \caption{Comparison of error field penetration threshold scalings between $\delta$ and normalized resonant flux. The perfect correlation (solid) line is surrounded by factor-of-2 lines (dashed) for reference.}
    \label{fig:threefigures}
\end{figure*}

While the dominant mode overlap has been demonstrated repeatedly to possess superior explanatory power compared to other error field metrics, it is not the only metric used presently in tokamak error field research. Two other metrics used are various versions of the total or vacuum resonant field at a rational surface. In \cite{Yang2021ParametricPlasmas}, overlap is shown to be superior to total and vacuum resonant fields for a single parameter scaling using density. However, the strength of these different metrics has never been explicitly compared for a multi-device, multi-parameter error field penetration threshold scaling. Using the data in this database, we produced scalings based on $\delta$, a vacuum resonant field metric, and a total resonant field metric. This is seen in Figures \ref{fig:comp_deltau}, \ref{fig:comp_b21}, and \ref{fig:comp_bv21}.

All three plots utilize the same forward least-squares regression outlined previously. As before, the Equation \ref{eq:delta_KDE} scaling has $R^2=0.66$ and an equivalent percent error of 36.6$\%$. The total resonant flux, $\delta B_{21} / B_{T}$, has $R^2=.15$, and an equivalent percent error of 54$\%$, and the vacuum resonant flux, $\delta B_{v21} / B_{T}$,  has $R^2=.29$ and an equivalent percent error of 55$\%$. The strong relative fit strength of $\delta$ provides further motivation for the adoption of the dominant mode overlap metric as the primary method for error field penetration scaling. 

The local resonant field is more numerically sensitive than $\delta$, especially to the quality of equilibrium reconstruction used. These local resonant metrics can produce highly nonphysical variance in prediction output \cite{logan_metrics_2025}. Since the dominant mode overlap has stronger correlations, it can produce much better quality fits, providing correspondingly higher predictive confidence. The improvement in $R^2$ from 0.15 to 0.66 is a significant increase in the amount of variance explained by the model, and the decrease in equivalent percent error from 54$\%$ to 36.6$\%$ is a substantial improvement in predictive uncertainty. This motivates the use of $\delta$ as the primary metric for error field penetration threshold scaling, as it provides a more accurate and reliable prediction of the thresholds observed in the database.

\section{\label{sec:UQ} Uncertainty Quantification} 
The uncertainty calculated and used throughout this work is produced from the least squares regression covariance estimate in log space \cite{montgomery2012introduction}. This estimates the uncertainty of each coefficient in the log scaling based on the residual scatter of the regression and the distribution and number of data points. Using this method, the uncertainty in the scalings represents variability from uncertainty in measurements, deviations from the scaling model assumed, and intrinsic variation. However, to consider the quality of the fit using a metric such as reduced $\chi^2$, we must provide an independent estimate of known measurement uncertainty.

There are many possible sources of uncertainty, for example it can result from differences in magnetic versus kinetic equilibrium reconstructions, as demonstrated in Reference \cite{logan_metrics_2025}, or from diagnostic measurements. One principal source of uncertainty can be estimated from the ramp speed of the 3D fields used to generate the locked modes. While this varies across devices, it can provide a useful upper bound. By taking a typical uncertainty of 10 ms in the time of the locked mode identification, we can propagate a percent error on the dominant mode overlap calculations. For example, in J-TEXT, locked mode experiments can have ramps from 0 to 5 kA on the order of 220 ms \cite{Mao2022StudyJ-TEXT}. The minimum 3D coil current among the J-TEXT shots that produced a locked mode was in shot 1073470, which was at 2.24 kA. Therefore, the ramp time to locked mode was $\approx 100 \text{ ms}$, so with 10 ms uncertainty, there is $10\%$ uncertainty in the lock time. Since coil currents are ramped roughly linearly, and the dominant mode overlap is linear in coil current, there is a $10\%$ uncertainty in dominant mode overlap. 

By using this 10\% value, we are able to calculate the reduced $\chi^2$ metric in an interpretable way. The closer the value is to one, the better the scaling prediction accounts for the observed error field penetration thresholds. By taking our $10\%$ uncertainty and projecting into log space, we calculate reduced chi-squared with a variance of $\ln(1+0.1^2)$. Assuming the same data uncertainty in the 2020 database, we can compare the predictive abilities of the old and new models by examining chi-squared values for scalings of the same type. 

For example, the ordinary least squares scaling from Reference \cite{logan_robustness_2020} has a reduced chi-squared value of 17.0, while the scaling in Equation \ref{eq:deltau_UQ} has one of 12.99. While both are greater than one, indicating a deficiency in predictive capacity, the decrease does indicate that the new scaling is statistically an improvement. Another useful metric for interpreting the performance of the scaling is the mean absolute log error. While the reduced chi-squared describes how well the scaling accounts for the thresholds observed in the dataset and is dependent on the scale of assumed uncertainty, the mean absolute log error is a scale-independent metric for multiplicative deviation. Mean absolute log error is a direct predictive metric that complements the reduced chi-squared, since the best model should minimize both the uncertainty in prediction and the statistical inconsistency of the predictions with the assumed uncertainty. The 2020 scaling has a mean absolute log error of 1.336, while the scaling from Equation \ref{eq:deltau_UQ} has one of 1.306. These can be interpreted as expected uncertainty on the threshold prediction of $\pm$ $33.6\%$ and $\pm$ $30.6\%$. Although the scaling in Equation \ref{eq:deltau_UQ} is a marginal improvement in prediction uncertainty, since it is also an improvement in reduced chi-squared, the scaling is more statistically reliable. However, the main improvement presented here is the consistent calculations of the overlap metric across all shots in the database.

Because the regression is performed in log space, $\delta = C \prod_i x_i^{\alpha_i}$, when the exponents, $\alpha_i$, are much larger than the values, $x_i$, standard error propagation is no longer valid. Instead, we use Monte-Carlo simulations of the exponent errors to produce prediction intervals represented by probability density functions, PDFs, and cumulative distribution functions, CDFs. Where the assumed uncertainty is propagated through the scaling to the exponents, it can be used to run Monte-Carlo simulations of error field penetration predictions, such as is done for ITER and SPARC \cite{pharr_error_2024,logan_sparc_2026}. This will be demonstrated in Section \ref{sec:ITER_proj} using the uncertainty on Equation \ref{eq:delta_KDE}.

\section{\label{sec:subsets} Impact of Subset Scalings}
Although the full-device scaling is generally the most helpful, it can be illustrative to downsample or remove a device or set of devices and rescale the database. This can be motivated in a variety of ways, including avoiding lower-fidelity equilibrium reconstruction, removing devices with important parameter space contributions, or creating minimal-device projections to future planned tokamaks. A selection of these subset scalings can be found in Table \ref{tab:subset_exps}. They are all of the form: 
\begin{equation}
    \delta = 10^{\alpha_C} I_p^{\alpha_{I}}n_e^{\alpha_{n}}R_0^{\alpha_{R}}B_T^{\alpha_{B}}\frac{\beta_n}{l_i}^{\alpha_{\beta}}
\end{equation}

\begin{table*}[t]
\centering
\caption{Scaling laws for $\delta$ with various machine subsets of the database, including $R^2$, $R^2$ of the scaling applied to the full dataset, condition number, reduced chi-squared, and mean absolute log error (MALE). Reduced $\chi^2$ is calculated based on an estimated 10\% uncertainty in the observed dominant mode overlap. Note that $R^2$ for Full OLS and Full WLS are not included since they already include the full datasets. The reduced $\chi^2$ for the WLS is also not included, since it relies on different underlying statistical assumptions than the other OLS scalings.}
\label{tab:subset_exps}

\setlength{\tabcolsep}{3pt}
\renewcommand{\arraystretch}{1.15}
\resizebox{\textwidth}{!}{
\begin{tabular}{l||cccccc||ccccc}
\hline
Data
& $C$
& $I_p$
& $n_e$
& $R_0$
& $B_T$
& $\beta_n/\ell_i$
& $R^2$
& $R^2_{\text{Full}}$
& Cond.\ No.
& Red. $\chi^2$
& MALE \\
\hline
Full OLS (Eqn. \ref{eq:deltau_UQ})
&-4.31 & -0.97 & 0.77 & 1.88 & 0.19 & 0.25 & 0.63 & - & 17.76 & 12.99 & 1.306 \\

Full WLS (Eqn. \ref{eq:delta_KDE})
&-4.26 & -1.01 & 0.56 & 1.57 & 0.30 & 0.13 & 0.66 & - & 17.76 & - & 1.366 \\

No J-TEXT
 &-4.22 & -1.00 & 0.62 & 1.63 & 0.27 & 0.25 & 0.64 & 0.59 & 18.84 & 13.41 & 1.317 \\

No DIII-D
 &-4.28 & -0.93 & 0.78 & 1.79 & 0.07 & 0.18 &  0.70 & 0.62 & 34.27 & 10.00 & 1.276 \\

No C-MOD
&-4.42 & -1.06 & 0.86 & 2.11 & 0.14 & 0.22 &  0.65 & 0.62 & 21.26 & 12.58 & 1.294 \\

JET, KSTAR, C-MOD
 &-4.04 & -1.05 & 0.44 & 1.29 & 0.08 & 0.10  & 0.80 & 0.24 & 44.36 & 7.71 & 1.237 \\
\hline
\end{tabular}}

\end{table*}

Looking at the table, we can consider as an example removing the J-TEXT data. Since J-TEXT is the only tokamak included without standard magnetic equilibria reconstruction, the TokaMaker code was used to recreate equilibria. Although TokaMaker is used widely and well validated, this represents a departure from the standardization striven for elsewhere across the database. Rescaling without J-TEXT provides a new expression for $\delta$ with highly similar parameters and fit metrics. This is a good indication that introducing the non-standard equilibria reconstruction does not have large-scale negative impacts on the database performance or predictive capability.

Typical error field studies in DIII-D \cite{Paz-Soldan2014,paz-soldan_spectral_2014} find a residual error field less than the empirical value, but still larger than the intrinsic error field thought to be in other machines. Although the present work increases the accuracy of the DIII-D error field treatment, it is still not a perfect representation. Therefore, DIII-D likely has a higher than typical uncertainty in error field modeling. As shown in the table, removing DIII-D does slightly increase the fit quality, with improvements in the subset $R^2$ compared to the fit in Equations \ref{eq:deltau_UQ} and \ref{eq:delta_KDE}. However, the $R^2_\text{full}$ of the full database using the scaling without DIII-D is lower than the full machine scaling, likely due to DIII-D's unique error fields. It causes a weakening of the major radius, and normalized pressure, but the density dependence and current stay essentially the same while the toroidal field dependence is reduced to nearly zero.

Also considered is a scaling removing C-MOD. Since C-MOD contains a majority of variation in the density parameter space, it is particularly interesting to examine the change in density dependence that C-MOD introduces. As seen in Table \ref{tab:subset_exps}, the plasma current dependency is similar, major radius is stronger, and toroidal field dependence is weaker. The density exponent increases significantly up to 0.86, so removing C-MOD leaves the scaling with a density dependence that is much closer to linear. While linear density dependence is aligned with past experiments \cite{Buttery2000ErrorJET}, more recent findings \cite{Yang2021ParametricPlasmas,Mao2022StudyJ-TEXT,Wang2018DensityEAST} agree with the non-linear density dependence as in Equations \ref{eq:deltau_UQ} and \ref{eq:delta_KDE}. This points to the importance of the high-density C-MOD data for the scaling to reproduce other nonlinear density scaling dependencies.

While there are countless other ways to downsample the devices, the last considered explicitly is trying to produce a minimal device scaling to ITER. Since DIII-D, EAST, and KSTAR all have similar operational space; we can choose one of the three, along with JET and C-MOD, to produce a scaling that encompasses much of the major radius range. A major limitation of this however, is that these scalings can have as few as 50 shots, compared to the full database of 127.

As seen in Table \ref{tab:subset_exps}, the JET, KSTAR, and C-MOD scaling seems promising with a very strong $R^2$ of 0.8, but when the scaling is used on the full dataset, the resulting $R^2$ is only 0.24. This is a hallmark of overfitting. The scalings with DIII-D and EAST performed similarly. The lack of parameter space coverage and small number of shots in these three-device scalings leads to overfitting and motivates the use of the full database for the most robust and generalizable scaling.

\begin{figure}[h]
\centering{}\includegraphics[width=1\linewidth]{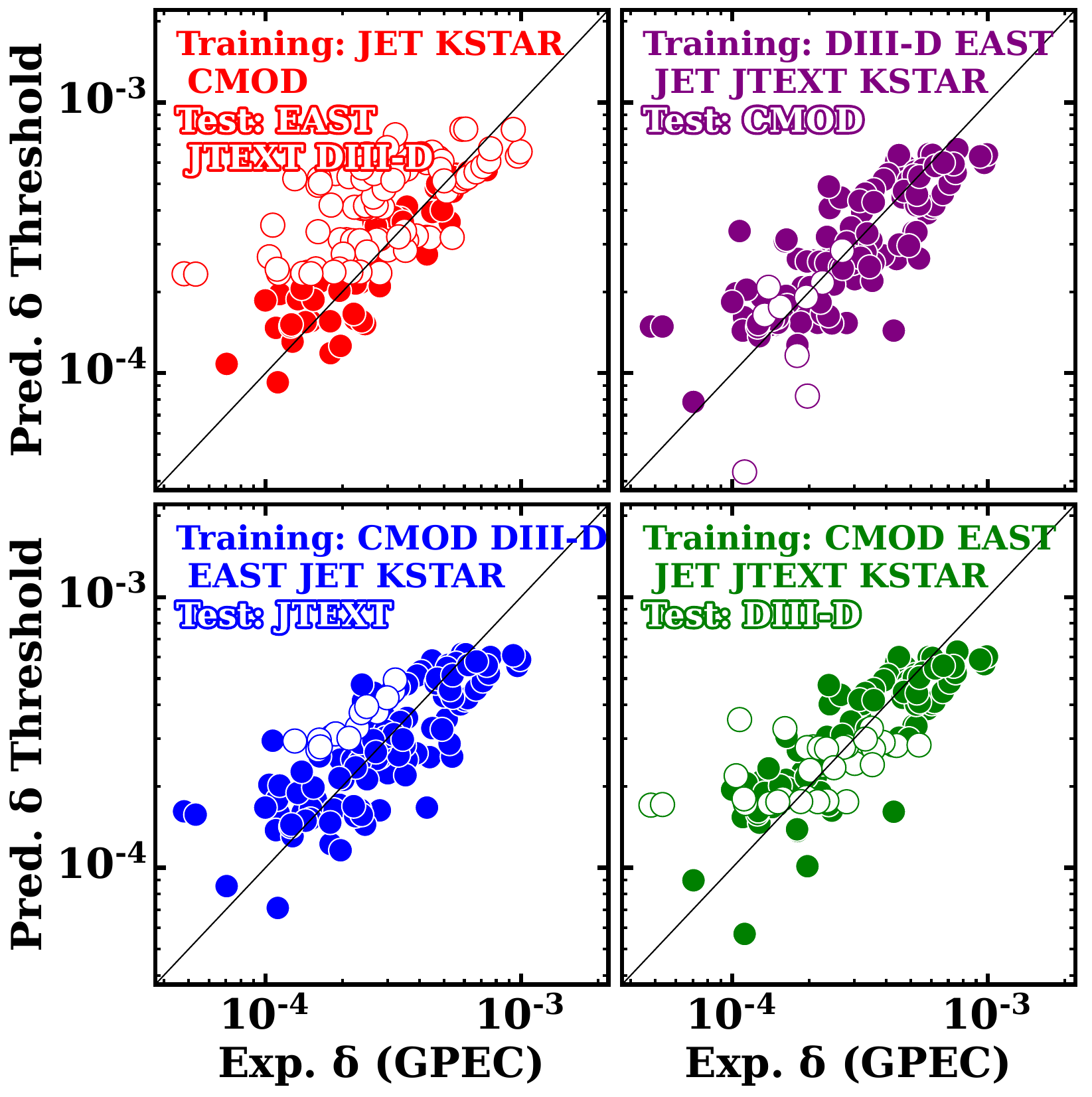}\protect\caption{Plots of actual vs predicted $\delta$ for a variety of device subsets. \label{fig:four_plot}}
\end{figure}

The dangers of this overfitting are visualized in Figure \ref{fig:four_plot}. Here, the whole database of all devices is divided up using various train test splits. In each subplot, the solid points are the training data fit to the subset scaling. The hollow points are the test data, from the remainder of the database, scaled using the subset scaling. The JET, KSTAR, and C-MOD scaling plot shows the high prediction uncertainty for the vast majority of the test data, most of which falls outside of the training data. In contrast, the subsets removing each of J-TEXT, DIII-D, and C-MOD have test data error that falls largely within that of the training dataset. These observations in the plot are confirmed by the fit metrics displayed in Table \ref{tab:subset_exps}. The single-device omissions demonstrate a higher degree of robustness to overfitting than in the three-device subset scalings. This implies any single device is not critical for a strong fit, although continuing to add additional devices and filling in the parameter space will likely continue to improve the scaling. 

\begin{figure}[h]
\centering{}\includegraphics[width=1\linewidth]{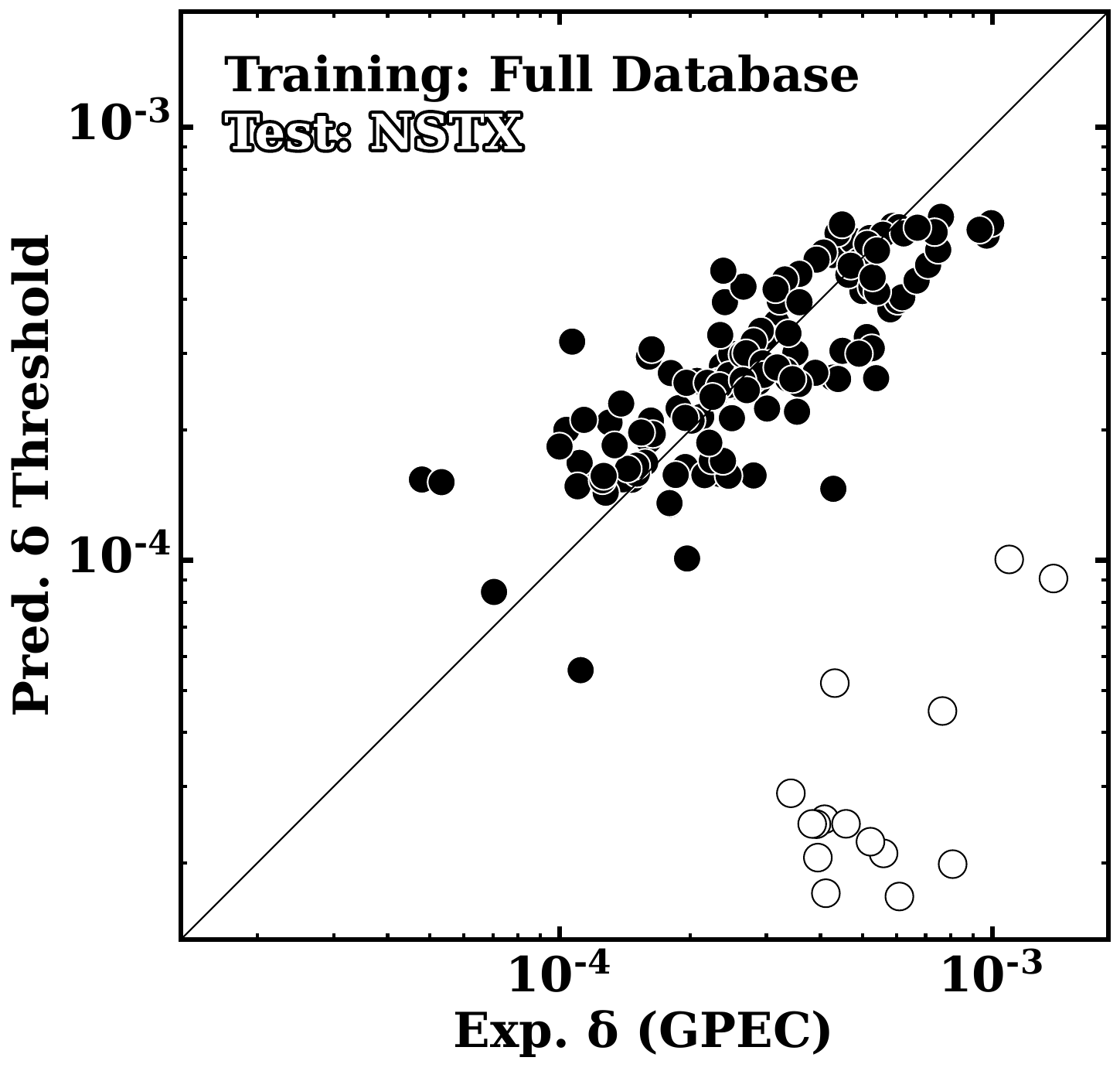}\protect\caption{NSTX predicted and actual $\delta$ using the new error field penetration threshold scaling.\label{fig:nstx_comp}}
\end{figure}

As noted previously, this scaling relies only on traditional, non-spherical tokamaks. NSTX was included in past scalings, but mostly due to the important part of the major radius parameter space it contributed to. With J-TEXT now included, NSTX differs from all other included devices by nature of being a spherical tokamak, without the benefit of adding a unique size point. 
Given the presented new scaling, we can apply it to the NSTX data used in previous work and assess its compatibility. Of course, there are many confounding factors that contribute to the relationship between NSTX and the scaling, but the main consideration here is its spherical nature. In Figure \ref{fig:nstx_comp}, NSTX is plotted alongside the full database using the scaling from Equation \ref{eq:deltau_UQ}. The $\delta$ predictions for NSTX are notably large underestimates. The actual values are roughly on the order of $10^{-4}$ but the predicted values are on the order of $10^{-5}$.

Unlike the devices in Figure \ref{fig:four_plot}, NSTX falls well outside the range of error arising from the training data. While the mechanism for NSTX being such an outlier is uncertain, one explanation is the much higher $q_{95}$ of NSTX as compared to the rest of the database. Since the q-profiles, and thus the rational surfaces differ from what is characteristically found in conventional tokamaks, and the dominant mode overlap is sensitive to the q-profile, this could be a key contribution to NSTX being a distinct outlier. Notably, this off-trend behavior is significantly more pronounced here than in previous work, since this is the first scaling to include explicit $B_T$ and $I_p$ dependence, and therefore the first to fully capture some amount of the $q_{95}$ dependence. Future work could shed more light on this issue through adding in other spherical tokamaks such as MAST-U, or by dedicated modeling of spherical tokamak error field penetration.

\section{\label{sec:ITER_proj} Projecting the Error Field Threshold to ITER}
This section compares the predicted ITER error field threshold values from past scalings to that of the scalings presented here. As has been done for ITER \cite{pharr_error_2024} and SPARC \cite{logan_sparc_2026}, the new scalings can be simulated using a Monte-Carlo method to arrive at error field penetration probabilities. By taking the scaling exponents and uncertainties from Equation \ref{eq:delta_KDE}, we produce a probability density of locking using $10^6$ samples.

\begin{figure}
    \centering
    \includegraphics[width=1\linewidth]{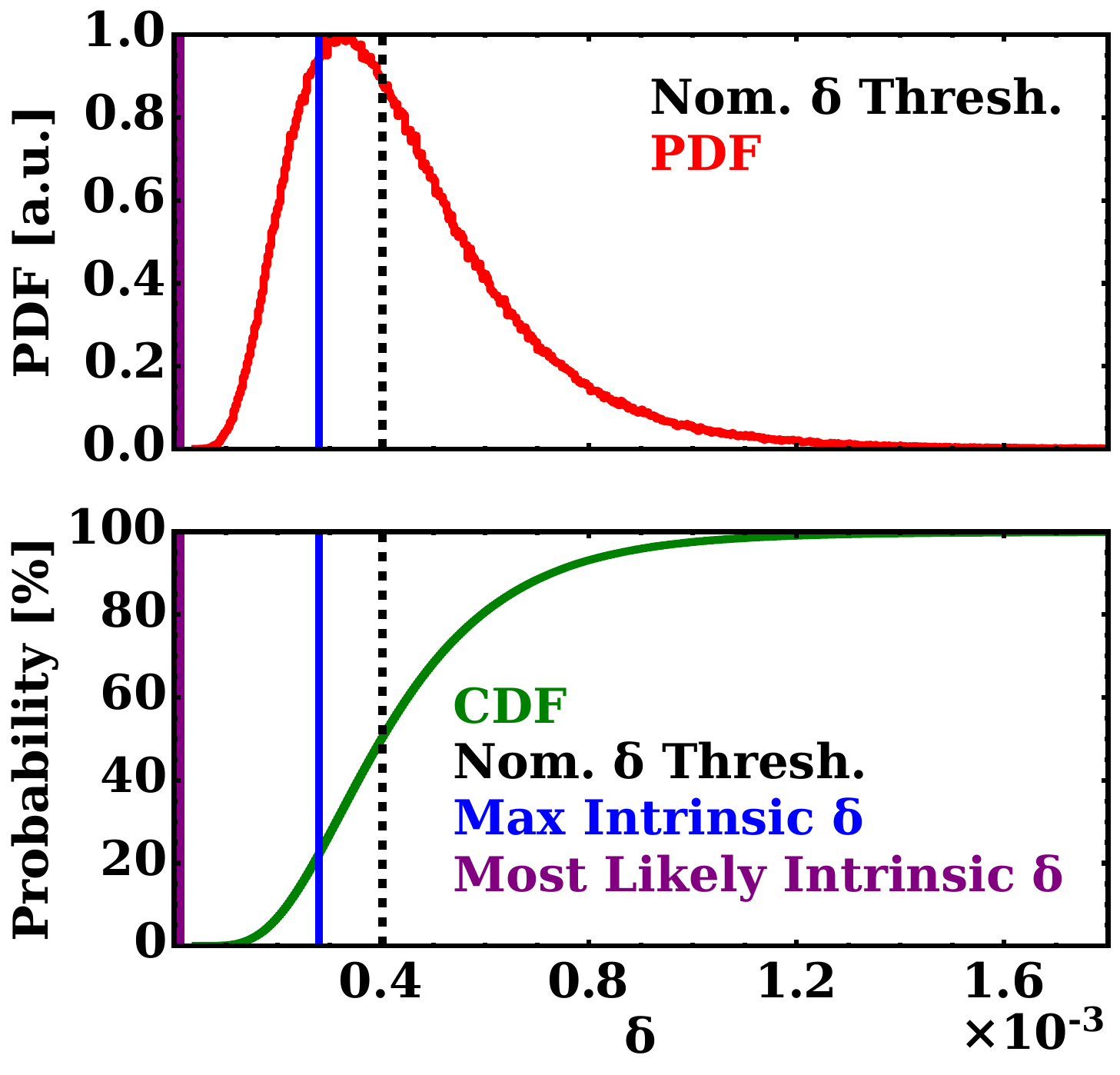}
    \caption{Monte-Carlo probability density function, PDF, (top) and cumulative distribution function, CDF, (bottom) for ITER's error field penetration threshold based on Equation \ref{eq:delta_KDE}. The intersection of the cumulative distribution function and the modeled intrinsic $\delta$ provides the probability that operation with a given set of plasma parameters will experience error field penetration. The black dotted line is the nominal $\delta$ threshold calculated from the scaling. The blue line is the maximum intrinsic $\delta$ for ITER and the purple is the most likely intrinsic $\delta$ for ITER, both taken from Reference \cite{pharr_error_2024}.}
    \label{fig:iter_mc}
\end{figure}

The top plot shows the probability density function, PDF, of the scaling law, based on $10^6$ random simulations of using the uncertainty on the exponents. By integrating the PDF, we arrive at the bottom plot. This is the cumulative distribution function, CDF, which gives the probability that the true value of $\delta$ is less than or equal to any specified value of $\delta$. By checking the probability of the CDF at the $\delta$ corresponding with a given intrinsic $\delta$, we arrive at the probability of locking for a given design point. Based on this analysis, the most likely value of $\delta$ for ITER's intrinsic error fields, $0.38\times10^{-4}$, has a probability of locking of 0.0001$\%$. The maximum possible value of $\delta$ for ITER's intrinsic error fields, $2.8\times10^{-4}$, has a locking probability of 22.3$\%$. These can both be seen plotted in the context of the PDF and CDF in Figure \ref{fig:iter_mc}. Given ITER's conservative design choices for error field mitigation, they continue to be well positioned to avoid locked modes. Note that a full risk assessment would consider the tail interactions between the probability density function of the intrinsic error field and the cumulative distribution function of the predicted error field penetration threshold.

\begin{table*}[t!]
\centering
\caption{Projections for ITER's error field penetration threshold. Uncertainty represents the symmetric average of the 25th and 75th percentile interval of the Monte-Carlo distribution of $\delta$.}
\label{tab:iter_projects_single}
{
\setlength{\tabcolsep}{8pt}
\renewcommand{\arraystretch}{1.5}
\begin{tabular}{lcc}
\hline
\noalign{\vskip 2.5pt}
Scaling Subset
& \shortstack{ITER Projection \\ ($\delta \times 10^4$)}
& \shortstack{Dimensional Threshold \\ $|\mathbf{v}^\dagger_1 \cdot \mathbf{\tilde{b}}^{x}|$ (Gauss)}
\\
\hline
\noalign{\vskip 2.5pt}
Previous Scalings \cite{logan_empirical_2020,logan_robustness_2020,bandyopadhyay_mhd_2025}
& $1.45 - 3.51$
& $7.69 - 18.63$
\\
\noalign{\vskip 2.5pt}
Full OLS Eqn. \ref{eq:deltau_UQ}
& $10.14 \pm 3.36$
& $53.70 \pm 17.80$
\\
\noalign{\vskip 2.5pt}
Full WLS Eqn. \ref{eq:delta_KDE}
& $4.02 \pm 1.29$
& $21.27 \pm 6.82$
\\
\noalign{\vskip 2.5pt}
No J-TEXT
& $5.92 \pm 2.30$
& $31.30 \pm 12.20$
\\
\noalign{\vskip 2.5pt}
No DIII-D
& $8.26 \pm 4.82$
& $43.80 \pm 25.50$
\\
\noalign{\vskip 2.5pt}
No C-MOD
& $10.43 \pm 4.99$
& $55.30 \pm 26.40$
\\
\noalign{\vskip 2.5pt}
JET, KSTAR, C-MOD
& $1.83 \pm 1.62$
& $9.70 \pm 8.60$
\\
\hline
\end{tabular}
}
\end{table*}

For a more complete picture, the predicted penetration threshold for ITER generated by each of the discussed scalings thus far is presented in Table \ref{tab:iter_projects_single}. These projections use the ITER "baseline scenario" specifications \cite{Poli2018ElectronScenario} including $B_{T} \approx 5.3$ T, $R_0 \approx 6.2$ m, $n_e \approx 9.8\times10^{19}$ m$^{-3}$ $\beta_n/l_i \approx 1.8$, and $I_p \approx 14.9$ MA. Each entry in the table is a projection using ITER parameters with errors that represent approximately 1 sigma of uncertainty in the value of $\delta$ based on the CDF of the scaling law. The dimensional threshold for each is the equivalent field strength in Gauss required to generate the given level of dominant mode overlap.

Since the range of critical locking thresholds for existing scalings is $\delta=1.45-3.51\times10^{-4}$ and the new scalings give $\delta=1.01\times10^{-3}$ and $\delta=4.02\times10^{-4}$, the new scaling predicts ITER will be more resilient to error fields than the past scalings. The other projections listed are all in the range of values encompassed by the previous and current scaling. All projections considered are considerably above both the most likely value, $0.38\times10^{-4}$,  and the expectation value, $\delta = 0.48\times10^{-4}$,  for ITER's intrinsic error fields \cite{pharr_error_2024}, indicating low risk of locking.

Projecting the scaling to other tokamaks falls into two categories: existing and planned. For existing tokamaks, different operating points can be entered into the scaling equation to assess the penetration threshold for a given shot. This can then be compared with GPEC modeling of similar shots to calculate the observed $\delta$. If the GPEC modeling shows a higher value than the predicted penetration threshold, the shot likely will experience mode activity and potentially a locked mode. For best accuracy, a model of the device's intrinsic error field should be included in the GPEC coupling calculations. For even more robust predictions, the Monte-Carlo modeling method of the error field penetration threshold can provide the probability that a given scenario will have penetration based on the uncertainty in the prediction threshold.  While this method benefits from being lightweight computationally, other more complex 3D magnetohydrodynamics modeling can often be a better choice when there is a need for high-fidelity on existing devices. The second category is for planned devices, such as ITER and SPARC. By using Monte-Carlo modeling of the tilts and shifts of the coils and other device parts, we can compare the probabilities of different intrinsic $\delta$ values with the Monte-Carlo probabilities of the threshold scaling $\delta$ value. By setting a tolerance for the probability of the device exceeding its threshold, we can prescribe engineering tolerances for the coils that determine the maximum allowable tilts and shifts that can be present without presenting an unacceptable risk of locked modes \cite{pharr_error_2024, logan_sparc_2026}. Here, the scaling threshold is essential, because higher fidelity modeling requires a much higher numerical precision of inputs than is physically possible in non-operational devices.

\section{\label{sec:Disc} Conclusions}
This paper presents a new error field penetration threshold scaling for operating and planned tokamaks. The overall best scalings considering dataset size, quality, and statistical metrics are the full device scalings found in Equations \ref{eq:deltau_UQ} and \ref{eq:delta_KDE}. The error field threshold database is now more self-consistent, relies only on traditional, non-spherical tokamaks, and uses only Ohmic and L-mode data. Additionally, the new scalings provide more physics intuition through the inclusion of explicit $I_p$ dependence. By including both the plasma current and the toroidal field, a more complete picture of device operating regimes and their implications for error fields is now possible. Increasing density continues to be one of the strongest predictors for protection from error fields, while increasing major radius and normalized pressure help to a smaller extent. 

Future work could include forming databases and producing scalings for other FPP relevant confinement regimes. Specifically, a SOC scaling from new experiments would complement the LOC scaling presented here. Low-rotational H-mode plasmas are important regimes for FPP and ITER extrapolation, where dedicated error field experiments are also needed. As negative triangularity (NT) develops into a promising ELM-free FPP scenario \cite{PazSoldan2024NegativeTriangularity,the_manta_collaboration_manta_2024,Freiberger2026CENTAUR,slendebroek_exploring_2026}, NT error field penetration scalings will be an important area of development. This dataset could be improved with the addition of new machines, especially JT-60 SA, since it would provide new data at large $R_0$, and ITER and SPARC once they start producing data. Since rotation is one of the most dominant features governing this physics, a positive future development would be to develop a scaling law based on all shots with rotation measurements. 

Both the OLS, Equation \ref{eq:deltau_UQ}, and the WLS, Equation \ref{eq:delta_KDE}, scalings predict an increase in the required level of error fields needed to cause locked modes on ITER. Their predictions are well above the most likely ITER overlap of $\delta = 0.38\times10^{-4}$ from Reference \cite{pharr_error_2024} (which per Section \ref{sec:distribution} is below 99\% of the penetration thresholds in the database). Upon consideration of all the scaling results, particularly the principal full device scaling, ITER remains highly likely to be free from locked modes during its baseline operating scenario. 

To conclude, the updated scalings introduced in this paper advance understanding of the error field penetration threshold and improve its prediction to critical operating scenarios for future devices. Including plasma current captures important behavior of the penetration threshold, and continuing to add data to the scaling is important for reducing collinearity and increasing confidence in the scaling predictions. The scalings can be used in conjunction with predictive modeling of intrinsic error field \cite{pharr_error_2024,logan_sparc_2026} for further design of new tokamaks, as well as the operation of existing ones, to predict and prevent the onset of disruptions from  mode-locking.

\section*{\label{sec:acknowledgements} Acknowledgments}

The authors acknowledge and thank all contributors to the joint error field activity MDC-19 within the ITPEA Topical Group on MHD, Disruptions and Control. In addition to the direct authors of this work, the original (Dr. Jong-Kyu Park, \href{mailto:jkpark@snu.ac.kr}{jkpark@snu.ac.kr}) and current (Dr. Nikolas Logan, \href{mailto:nikolas.logan@columbia.edu}{nikolas.logan@columbia.edu}) lead of MDC-19 would like to thank Drs. Y. Sun (EAST), V. Igochine (AUG), M. Maraschek (AUG), S. Munaretto (NSTX-U), Y. In (KSTAR), and T. Markovic (COMPASS) for their empirical contributions to collaborative error field research and helpful discussions.

Data analysis and modeling in this work was facilitated by the OMFIT integrated modeling framework \cite{OMFIT2015}. 

This work has been carried out within the framework of the EUROfusion Consortium, funded by the European Union via the Euratom Research and Training Programme (Grant Agreement No 101052200-EUROfusion). Views and opinions expressed are however those of the author(s) only and do not necessarily reflect those of the European Union or the European Commission. Neither the European Union nor the European Commission can be held responsible for them.

This work was supported by the U.S. Department of Energy Office of Science Office of Fusion Energy Sciences using the DIII-D National Fusion Facility and Alcator C-Mod, both DOE Office of Science user facilities,
and under DOE Awards DE-FC02-04ER54698, DE-FC02-99ER54512, DE-SC0022270, and DE-SC0021968.

This report was prepared as an account of work sponsored by an agency of the United States Government. Neither the United States Government nor any agency thereof, nor any of their employees, makes any warranty, express or implied, or assumes any legal liability or responsibility for the accuracy, completeness, or usefulness of any information, apparatus, product, or process disclosed, or represents that its use would not infringe privately owned rights. Reference herein to any specific commercial product, process, or service by trade name, trademark, manufacturer, or otherwise does not necessarily constitute or imply its endorsement, recommendation, or favoring by the United States Government or any agency thereof. The views and opinions of authors expressed herein do not necessarily state or reflect those of the United States Government or any agency thereof.

\printbibliography

\end{document}